\documentclass[12pt,a4paper]{article} 
\usepackage{epsfig}
\usepackage{color}
\usepackage{array,multirow,graphicx}
\usepackage{float}
\usepackage{amsmath}
\usepackage{hyperref}
\usepackage{dcolumn}
\usepackage[position=top]{subfig}
\usepackage{float}

\newcommand{\tildd}{{\raise.17ex\hbox{$\scriptstyle\sim$}}}

\usepackage{appendix}
\usepackage{authblk}

\title{Analysis of the C$_2$ (d$^3\Pi_g$-a$^3\Pi_u$) Swan bands as a thermometric probe in CO$_2$ microwave plasmas}

\author[1]{Emile Carbone, Federico D'Isa, Ante Hecimovic, Ursel Fantz}
\affil[1]{Max Planck Institute for Plasma Physics, Boltzmannstr. 2, D-85748 Garching, Germany\\
Corresponding author: emile.carbone@ipp.mpg.de}

\date{} 

\begin{document}
\maketitle





\begin{abstract}
The optical emission spectra of high pressure CO$_2$ microwave plasmas are usually dominated by the C$_2$ Swan bands. In this paper, the use of the C$_2$ Swan bands for estimating the gas temperature in CO$_2$ microwave plasmas is assessed. State by state fitting is employed to check the correctness of assuming a Boltzmann distribution for the rotational and vibrational distribution functions and, within statistical and systematic uncertainties, the C$_2$ Swan band can be fitted accurately with a single temperature for rotational and vibrational levels. The processes leading to the production of the C$_2$ molecule and particularly its d$^3\Pi_g$ state are briefly reviewed as well as collisional relaxation times of the latter. It is concluded that its rotational temperature can be associated to the gas temperature of the CO$_2$ microwave plasma and the results are moreover cross-checked by adding a small amount of N$_2$ in the discharge and measuring the CN violet band system. The 2.45~GHz plasma source is analyzed in the pressure range 180-925~mbar, for input microwave powers ranging from 0.9 - 3 kW and with gas flow rates of 5-100~L/min. An intense C$_2$ Swan bands emission spectrum can be measured only when the plasma is operated in contracted regime. A unique temperature of about 6000 $\pm$ 500 K is obtained for all investigated conditions.  A spectroscopic database is constructed using the recent compilation and calculations by Brooke et al. \cite{BROOKE201311} of the line strengths and molecular constants for the C$_2$ (d$^3\Pi_g$-a$^3\Pi_u$) Swan bands system and made available as Supplementary Material in a format compatible with the open source MassiveOES software.

\end{abstract}

Published as: \textit{Plasma Sources Sciences and Technology (2020), \textbf{29}, 055003}

\section{Introduction}
\label{intro}

Many strategies are being currently studied for the recycling and valorization of CO$_2$, a potent greenhouse gas in the atmosphere \cite{ofelia2014co2}. Production of liquid fuels from renewable energies is one of the pathways for re-utilization of CO$_2$. The first step for converting CO$_2$ into added value chemicals is its conversion into carbon monoxide CO \cite{Lebouvier2013, SnoeckxCSR2017}. For practical applications, it is desirable to operate with high gas flows near atmospheric pressure (or above) in order to avoid bulky components, reduce costs and save energy for the compression steps \cite{vanRooij2018PPCF}. Previous studies have indicated that non-equilibrium microwave plasmas are particularly suitable for high energy efficiencies $\eta$ conversion of CO$_2$ into CO with $\eta\leq$ 80\% and can operate at high pressure and with high flow rates \cite{Fridman2008}. CO$_2$ based plasmas have applications in fields such as lasers \cite{Sobolev1967}, plasma polymers treatment \cite{FAVIA19981102} and also have potential in plasma medicine \cite{Carbone2018}. The proposed mechanism for explaining high energy efficiencies for CO$_2$ conversion into CO is via the so-called vibrational ladder climbing mechanism. Relatively low energy electrons excite preferentially the assymetric vibrational modes of the CO$_2$ ground state CO$_2 (00 i)$ molecule and vibrational energy exchange between CO$_2$ molecules leads ultimately to their decomposition \cite{Rusanov1981}. To favor this mechanism the key ingredient is to keep the plasma out of equilibrium with the gas temperature $T_g$ much smaller than the effective electron temperature $T_e$ \cite{Kosak2014, Rooij2015taming, Capitelli2017vib}. Low gas temperatures (i.e. $T_g\ll 3000$ K) are required in order to avoid too many losses of CO$_2 (00 i)$ states by collisions with the background gas leading to gas heating by vibrational to translation relaxation processes \cite{treanor1968vibrational, azizov1983nonequilibrium}. For high energy efficiencies, calculations also indicate that formed O atoms need to react with highly vibrationally excited CO$_2$ molecules and to produce another CO molecule and a O$_2$ molecule \cite{Bongers2017}. On the other hand, at high gas temperatures (i.e. T$_{gas}\geq 3000$ K), thermal dissociation is the dominant mechanism for CO$_2$ dissociation with high conversion rates and is limited to energy efficiencies up to $\sim 50$ \% \cite{denHarder2017, Bekerom2018}.

\begin{figure}[ht]
\centering
\includegraphics[width=0.95\textwidth]{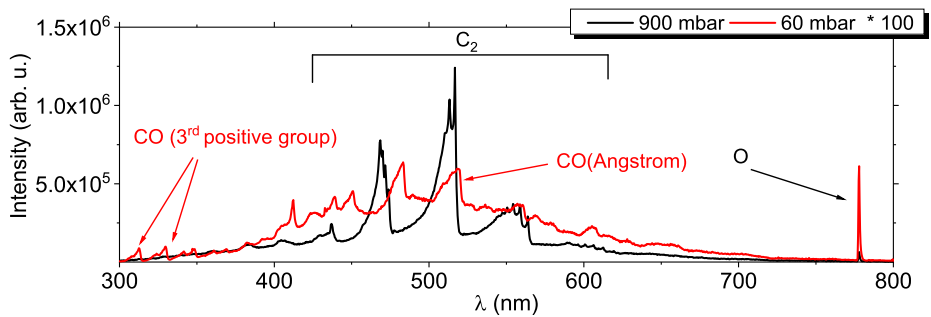}
\caption{Overview spectra using a low resolution spectrometer (instrumental profile of about 1.2 nm) of the plasma emission at low and high pressures. The high pressure spectrum is characteristic for all the conditions of the present study where the plasma is in contracted mode which occurs for power $\geq 900~W$ at pressures above 120 mbar. Note that the spectra are not calibrated for the sensitivity response of the spectrometer and optics.}
\label{fig:SPEC_overview}
\end{figure}

For studying the mechanisms limiting the plasma energy efficiency of the plasma, it is critical to have tools for assessing the degree of non-equilibrium in the plasma phase and notably its gas temperature. In high pressure (i.e. sub- up to atmospheric pressure) microwave CO$_2$ plasmas, one of the most dominant emission feature is the C$_2$ (d$^3\Pi_g$-a$^3\Pi_u$) Swan bands system. In figure \ref{fig:SPEC_overview} an overview spectrum of a pure CO$_2$ microwave plasma (see section \ref{setup} for details on the plasma setup) is shown. In addition to a weak continuum emission\footnote{This continuum emission which is the dominant emission feature in the effluent of the plasma and at its edges corresponds to chemiluminescence of CO$_2$* formed by recombination of O+CO species \cite{GUPTA20113131, SAMANIEGO1995}.} and typical atomic oxygen and carbon lines\footnote{Carbon atom lines are only  observed below 300 nm and above 900 nm and are not shown in figure \ref{fig:SPEC_overview}.}, the C$_2$ Swan bands are the dominant molecular emission structure in the UV and visible range. Such spectrum is typical and has been measured by several authors \cite{Bongers2017, Babou2008, hong2011generation, LeQuang2012, Spencer2013MWcat, UHM2016, Mitsingas2016, SunHojoong2017, Chun2017reforming}. To obtain reliable and accurate information about the plasma, it is then critical to have tools for evaluating the emission spectrum of the C$_2$ molecule and assess the accuracy of the results. Usually an equivalence between the rotational temperature of emitting species and the gas temperature is assumed but this hypothesis needs to be assessed. Also, in certain cases non-Boltzmann distributed rotational distribution function are measured experimentally \cite{Bruggeman2014RotT}. In combustion studies, it has been observed that the C$_2$ Swan band can exhibit overpopulation of the high vibrational levels. The latter emission spectrum is known as the \lq\lq high pressure bands\rq\rq of the C$_2$ system \cite{pearse1976identification}. This makes the analysis of C$_2$ spectra more complicated and can induce systematic errors in the determination of the rotational temperature (T$_{rot}$) of the C$_2$ (d$^3\Pi_g$) state if the vibrational levels are assumed to take a Boltzmann distribution function. For CO$_2$ plasmas, an evaluation of potential deviations of C$_2$ Swan bands spectra from Boltzmann equilibrium and the assessment of $T_{rot}$ as a measure of the gas temperature have not been reported in the literature yet.


 In this paper we present the spectral fitting analysis of the C$_2$ Swan band using a newly implemented spectroscopic database in the open source MassiveOES software \cite{Vorac2017, Vorac2017b}. The database is constructed using the recent compilation and calculations by Brooke et al. \cite{BROOKE201311} of the line strengths and molecular constants for the C$_2$ (d$^3\Pi_g$-a$^3\Pi_u$) Swan band system. C$_2$ Swan bands spectra measured on a 2.45 GHz CO$_2$ microwave plasma source are reported and carefully analyzed. By adding small amount of nitrogen in the plasma, the obtained temperatures from the Swan band fitting are compared with the one obtained by fitting the CN violet band system. Uncertainties related to the method used for fitting the C$_2$ spectrum, potential deviations from a Boltzmann distribution for the rotational and vibrational distribution functions and the assumption of fitting the spectra with a single or two temperatures (i.e. rotational and vibrational temperatures) are discussed and compared with a state-by-state fitting of the experimental spectra. Finally, currrent knowledge about rotational and vibrational relaxation processes of the C$_2$ ($d^3\Pi_g$) state is presented in section \ref{C2relax} and the use of T$_{rot}$ for measuring the gas temperature of CO$_2$ microwave plasmas is discussed.

\section{Experimental setup}
\label{setup}

\begin{figure}[ht]
\centering
\includegraphics[width=0.75\textwidth]{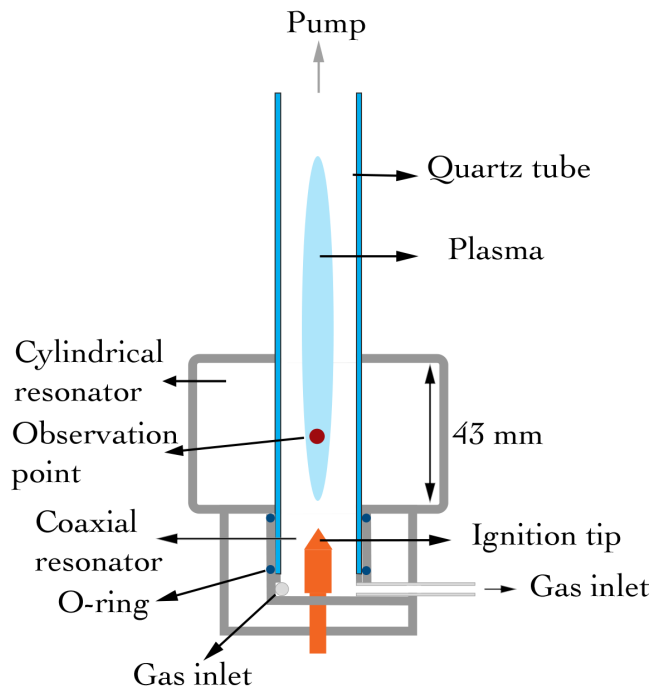}
\caption{Schematic representation of the 2.45~GHz plasma source. The gases are tangentially injected at the bottom of the coaxial cavity (only two of the 4-inlets are represented.}
\label{fig:drawing} 
\end{figure}

A microwave device which can ignite an atmospheric pressure plasma torch using only microwave power without the use of additional igniters \cite{leins2015ignite} is studied. A modified version of this device has been built so that it can be operated from low (\tildd 10 mbar) up to atmospheric pressure in molecular gases such as N$_2$ and CO$_2$. Stable operation of the plasma is obtained both in continuous and pulsed regime in a large parameters range. The microwaves are coupled via a WR340 waveguide (TE$_{01}$ mode) into a cylindrical cavity with low Q-factor where the plasma is sustained inside a quartz tube (see figure \ref{fig:drawing} for more details). The quartz tube has inner/outer diameters of 26/30~mm.

For the ignition of the plasma, a coaxial cavity made of a waveguide to cylindrical resonator with high Q-factor and an inner conductor with conical shape is integrated at the bottom of the quartz tube. The coaxial cavity serves also for the injection of the gases using a 4-inlets tangential gas injection system. The coaxial conical pin vertical position can be adjusted for selecting different modes and enhance the electric field at its tip to ignite the plasma. The plasma can be operated in the power range of 600~W-3~kW, with gas flow rates between 5-100 L/min and pressures from 10 up to 960 mbar.

The high resolution measurements were acquired using a Czerny-Turner 1~m focal length SPEX1000 spectrometer with an entrance slit of about 30~$\mu$m and a dispersion grating of 1800 g/mm. An Andor iDus 420 CCD camera was used to record the spectra. The spectral resolution in first order is 20~pm at 550~nm. The wavelength calibration of the spectrometer was done using a mercury lamp.  The light is collected using a system of two irises and a lens. The lens focuses the light into an optical fiber system that is used for coupling the light to the entrance slit of the spectrometer. The irises system acts as a collimator and a spatial resolution of 1~mm is obtained. The collecting optics, filters and the spectrometer have been absolute calibrated between 400 nm to 850 nm with an Ulbricht sphere to obtain the absolute photon flux from the plasma. A longpass filter at 400~nm (FEL0400 from Thorlabs) is used to suppress any contribution from second order emission overlapping with the Swan band emission. For the CN violet system measurements, a 280~nm longpass filter (FEL280 from Thorlabs) is used. Additional low-resolution survey spectra are acquired using an Ocean Optics S2000 spectrometer with 1.2 nm resolution.


\section{Sources of C$_2$ and of the C$_2$ (d$^3\Pi_g$) state}
\label{C2kin}

The primary building blocks of C$_2$ are C atoms and it is therefore of interest to look first at sources of C atoms. No direct production channel from the CO$_2$ molecule is known by electron impact dissociation \cite{mcconkey2008electron, Grofulovic2016}. Electron impact dissociation processes of CO$_2$ lead to O and CO fragments either as neutral (ground or excited states). Electron dissociative attachment to CO$_2$ peaks at low electron energies $<$ 10 eV and only produces CO \cite{Nag2015} and O$^-$. Electron impact ionization of CO$_2$ leading to formation of C$^+$ has a threshold cross section of 27.8~eV (compared to 13.8~eV for CO$_2^+$) according to Itikawa \cite{Itikawa2002CO2} and can therefore be considered as negligible while considering typical electron energy distribution function in CO$_2$ plasmas \cite{Pietanza2016CO2eedf}. Dissociative recombination of CO$_2^+$ lead mostly to CO and O products and a small fraction of C atoms \cite{FLORESCUMITCHELL2006277}. However, CO$_2^+$ is lost by charge exchange and clustering reactions and is not a dominant ionic species in the plasma \cite{Carbone2017ispc}. CO$_2$ species cannot then be considered as a primary source of C atoms and  CO can be regarded as its primary source.

\begin{figure}[ht]
\centering
\includegraphics[width=0.75\textwidth]{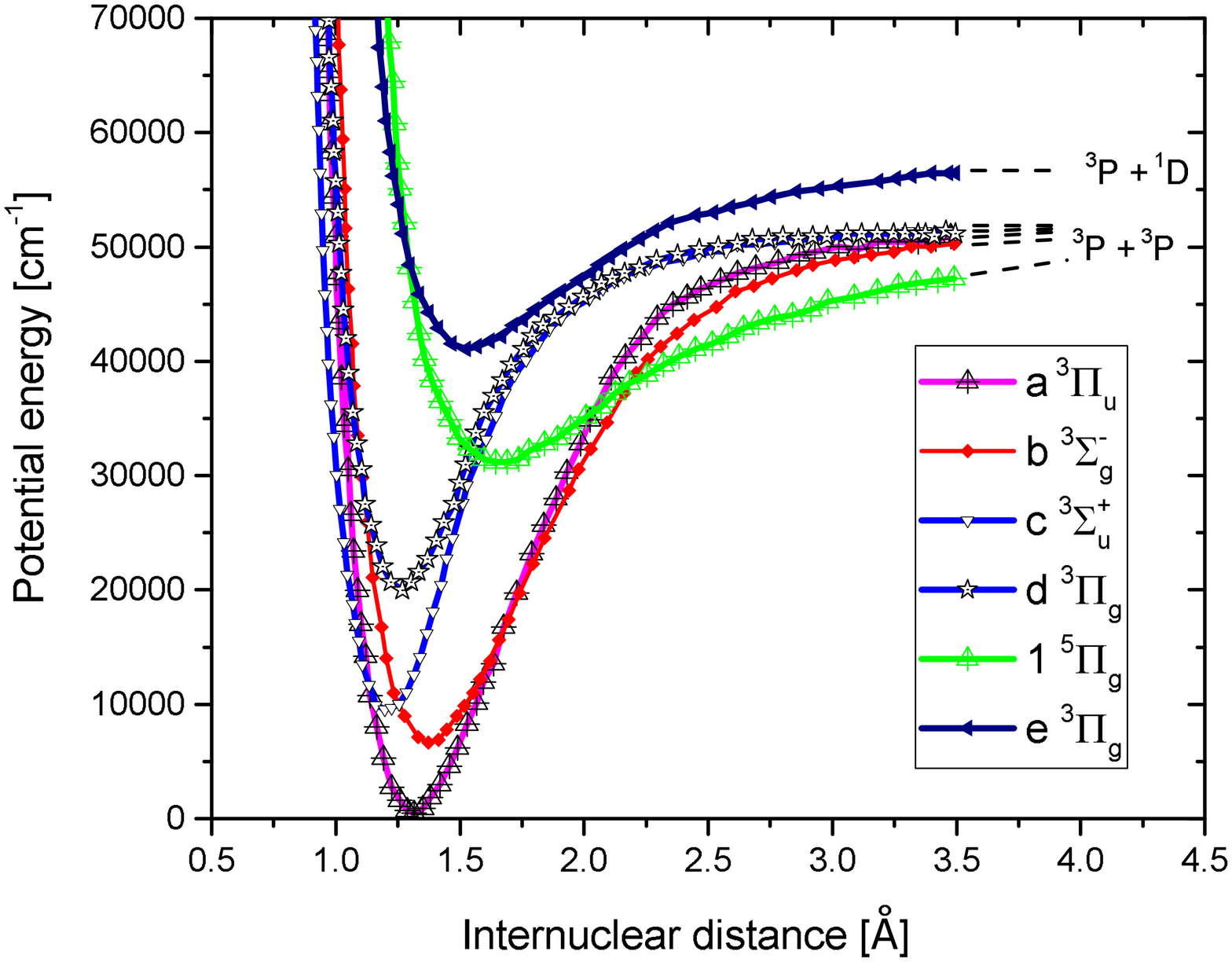}
\caption{Schematic representation of the potential energy curves of the lower triplet electronic states of the C$_2$ molecule as well as of the $1^5\Pi_g$ state and their dissociation limit. For more information see ref. \cite{martin1992c2}. }
\label{C2potentialcurve} 
\end{figure}

Electron impact dissociation and recombination processes of CO and CO$^+$ have been extensively reviewed (see for instance \cite{Pietanza2017CO, FLORESCUMITCHELL2006277}) and we will focus here on neutral processes leading to the formation of C atoms or even directly to C$_2$ molecules. A reason for focusing on these processes is also the knowledge that, in high pressure discharges with high degree of vibrational excitation, neutral processes are expected to play an important role for molecules dissociation \cite{Fridman2008}. Moreover, in section \ref{sec:results} are reported high gas temperatures for which thermal (i.e. neutral) processes dominates dissociation dynamics \cite{Bekerom2018}.

To understand the formation of the C$_2$ d$^3\Pi_g$ state, as it will be discussed in the following, not only source terms of C atoms but also of the C$_2$O should be considered. Vibrationally excited CO molecules lead to the formation of carbon atoms via thermal decomposition \cite{KUSTOVA2001638} or via the so-called Boudouard mechanism \cite{Serdyuchenko2009}

\begin{equation}
\label{boudouardClassic}
CO (X^1\Sigma_g, v) + CO (X^1\Sigma_g, w) \rightarrow CO_2 + C
\end{equation}

where $v$ and $w$ are vibrational quantum numbers.\footnote{Alternatively of having two CO molecules in their ground state $X^1\Sigma_g$ in the reaction \ref{boudouardClassic}, one of the species reacting can be a CO(a$^3\Pi$,v) excited state formed from energy exchange between two CO (X$^1\Sigma_g$, v) molecules \cite{Barreto2017}.} This process is very efficient and researchers have reported that discharges into carbon monoxide even at pressures of a few Torr lead to deposition of solid carbon on the walls and a decrease of the pressure in the closed cell \cite{kini1977investigation} . A competing process to reaction \ref{boudouardClassic} is the formation of a C$_2$O molecule \cite{Barreto2017, WILLIS1968}

\begin{equation}
\label{C2Oform}
CO (X^1\Sigma_g, v) + CO (X^1\Sigma_g, w) \rightarrow C_2O + O.
\end{equation}

Such processes are efficient only when significant vibrational excitation of CO molecules is present and/or high gas temperatures. This is the case in the present study with reported temperatures of 6000~K (see section \ref{sec:results}) where CO$_2$ is thermally dissociated, and production of C atoms happens in the gas phase\cite{Fridman2008}. 

The formation of the C$_2$ molecule from the three lowest electronic configurations of the carbon atom which are the 2s$^2$2p$^2$ ($^3$P$_J$, $^1$D$_2$ and $^1$S$_0$) states, where the ground state is the $^3$P$_0$ state, was discussed by Ballik and Ramsay \cite{ballik1963extension} while considering the potential energy curves of C$_2$. In figure \ref{C2potentialcurve} are shown the lower triplet electronic levels of the C$_2$ molecule. For a recent overview of the potential energy curves of the C$_2$ molecule we refer to Martin \cite{martin1992c2}. The d$^3\Pi_g$ state \footnote{In Ballik and Ramsay this state is denoted as the A$^3\Pi_g$ state which is a common notation found for the Swan band in the literature with the lower state of the transition defined as X$^3\Pi_u$.} is formed via

\begin{equation}
\label{CCassoc}
C (^3P_J) + C(^3P_J) + M\rightarrow C_2(d^3\Pi_g) + M
\end{equation}

where M stands for a third body that stabilizes the transition state. They noted that there is an avoided crossing with the C$_2$(e$^3\Pi_g$) state that dissociate into C ($^3$P) + C($^1$D) and so that the C($^1$D) does not recombine into the C$_2$(d$^3\Pi_g$) state. The C$_2$ Swan band can then potentially be used as marker for the recombination of ground state C atoms. 

In high pressure discharges, a preferential emission of the C$_2 (d^3\Pi_g, v=6)$ state can be observed which is known as the \lq\lq C$_2$ high pressure band\rq\rq. C$_2$O molecule was proposed by Kunz et al. \cite{Kunz1967} for explaining the preferential emission of the C$_2 (d^3\Pi_g, v=6)$ state. The C$_2$O molecule\footnote{Note that in addition to reaction \ref{C2Oform}, the C$_2$O molecule can be formed from the recombination of a C atom with CO via

$$C + CO + M\rightarrow C_2O + M.$$

The formation of C$_2$O in pulsed radiolysis of CO was shown first by Willis and Devillers \cite{WILLIS1968} and they showed that adding CO$_2$ did not affect the kinetics of C$_2$O. Kinetics of this molecule discussed from kinetic analysis of pure CO discharges can then probably be extended to mixtures of CO/CO$_2$. } reacts with a C atom and forms an electronically excited C$_2$ molecule \cite{Wallaart1995}

\begin{equation}
\label{C2Oswan}
C_2O + C \rightarrow C_2^* + CO.
\end{equation}

Naegeli and Palmer proposed that the initial electronic state in reaction \ref{C2Oswan} is the C$_2$(b$^3\Sigma_g^-)$ state that crosses to the C$_2$ (d$^3\Pi_g$) state at v=6 and leads to the high pressure C$_2$ band \cite{NAEGELI1968}. Through a kinetic study of a CO plasma afterglow, Gosse et al.\cite{gosse1972} supported the hypothesis that C$_2$(d$^3\Pi_g$) state is formed via reaction \ref{C2Oswan}. Brewer et al \cite{Brewer1962} stated that formation of the high pressure band from a predissociation mechanism is unlikely and suggested that a resonant transfer from another molecule happens. However, Caubet and Dorthe \cite{Caubet1994origin}, C$_2$O may also be a source of the $^5\Pi_g$ state via

\begin{equation}
\label{C2OCaubet}
C_2O (X^3\Sigma^+) + C (^3P) \rightarrow C_2(^5\Pi_g, v=0) + CO (X^1\Sigma^+).
\end{equation}

They argued that based on the correlation diagram for the reactants and products of the reaction C$_2$O + C that the $^5\Pi_g$ state of C$_2$ is correlated to the ground state of the reactants while the d$^3\Pi_g$ state is not.
We note that previously, because of the absence of observation of C$_2$O by absorption, Kini and Savadatti \cite{kini1977investigation} proposed the following mechanism for the formation of the high pressure C$_2 (d^3\Pi_g, v=6)$ Swan band

\begin{equation}
\label{C3meca}
C_2(a^3\Pi_u) + C_3(m)\rightarrow  C_2 (d^3\Pi_g, v=6) + C_3
\end{equation}

where the C$_3$(m) molecule is a postulated metastable species with an energy of 3.7 eV. Little and Browne \cite{little1987origin} argued against the possibility of the C$_3$ as intermediate species for the formation of the high pressure Swan band. Indeed, the C$_2$ high pressure band is observed in a large range of experimental conditions where C$_3$ formation is not favored and they also advanced a lack of evidence for a metastable state that should have an energy above the $\tilde{A}^1\Pi_u$ state which has 3.1 eV internal energy and is responsible for the C$_3$ comet band ($\tilde{A}^1\Pi_u$-$\tilde{X}^1\Sigma_g^+$ transition). Recent spectroscopic investigation and calculations of the energy levels of the C$_3$ molecule indicate that the lowest triplet state C$_3$($\tilde{a}^3\Pi_u$) has an energy of about 2.1 eV and the $\tilde{b}^3\Pi_g$ an energy of 2.9 eV \cite{Sasada1991, Terentyev2004} which are well below the 3.7 eV energy required for the process \ref{C3meca}. The C$_3$ molecule can therefore be ruled out as source of the C$_2$ high pressure band.

Little and Browne \cite{little1987origin} proposed a general mechanism for the formation of the high pressure Swan band via the following three steps

$$C (^3P) + C(^3P) + M\rightarrow C_2(^5\Pi_g, v) + M$$
$$C_2(^5\Pi_g, v) + M \rightarrow C_2(^5\Pi_g, v=0) + M$$
$$C_2(^5\Pi_g, v=0) + M\rightarrow C_2 (d^3\Pi_g, v=6) + M$$

where the quintet $^5\Pi_g$ is a metastable state for which a crossing exists between the v=6 level of the $d^3\Pi_g$ state and the v=0 level of the quintet state. Kirby and Liu calculated its energy to lies $3.85\pm0.34$ eV above the ground state of C$_2$ \cite{Kirby1979} while Little and Browne estimated a value of 3.79~eV. Such mechanism essentially allow to explain previous observation that the high pressure band is rotationally colder than the \lq\lq normal\rq\rq~ Swan band emission that can be observed in the same discharge \cite{faust1981time}. Also, high resolution analysis of the C$_2 (d^3\Pi_g, v=6)$ low rotation energy levels shows that both odd and even rotational numbers are perturbed with increase of intensity \cite{Meinel1968} which indicates that a $\Sigma$ state is not involved in the perturbation but a state with $\Lambda\neq 0$ fills the lower rotational levels of the  C$_2 (d^3\Pi_g, v=6)$ state \cite{Caubet1994origin}. A detailed analysis shows in fact that the $b^3\Sigma_g^-$ state perturb the J=19 and 21 of the v=6 upper state of the Swan band but lead to a decrease of the intensity of these lines \cite{little1987origin}. It can be concluded that the $b^3\Sigma_g^-$ state is not at the origin of the high pressure band and that the quintet $^5\Pi_g$ state is likely its source.


The proposal that C atoms recombine via the quintet $^5\Pi_g$ state which then populate the C$_2 (d^3\Pi_g, v=6)$ state (i.e. to generate the C$_2$ high pressure band) does not exclude the direct recombination of C atoms into the $d^3\Pi_g$ state. One should see them as processes that happen simultaneously with different probabilities. Looking at the energy potential curves of the C$_2$ molecules, one can see that the X$^1\Sigma_g^+$, A$^1\Pi_u$, B$^1\Delta_g$, B'$^1\Sigma_g^+$, C$^1\Pi_g$, a$^3\Pi_u$, $b^3\Sigma_g^-$, $c^3\Sigma_u^+$, $d^3\Pi_g$ and $^5\Pi_g$ all have as dissociation limit two C($^3$P) atoms \cite{martin1992c2}. The branching ratios for the recombination into these different electronic states are however not known although it appears from experiments discussed before that the $d^3\Pi_g$ and $^5\Pi_g$ are the ones first (and mainly) populated. 

In conclusion, we can say that there are several concurring production mechanisms of the C$_2$ (d$^3\Pi_g$) state either from recombination of C atoms or, in the case of CO containing plasmas, via the C$_2$O molecule. The presence or absence of some of these mecanisms for which branching can unfortunately hardly be estimated will affect particularly the nascent population distribution of vibrational states of the C$_2$ (d$^3\Pi_g$) state.

\section{Molecular spectra calculations}
\label{MolSpecAnal}

In order to reproduce experimental spectra, it is necessary to first model the population distribution of states and, using molecular constants and transitions probabilities, one can simulate a spectrum knowing the instrumental profile of the apparatus. The partition functions of rovibrational distributions under the assumption of a Maxwell-Boltzmann distribution used to generate so-called Boltzmann plots are given in section \ref{sec:thermMB}. The molecular constants used for calculating both the C$_2$ Swan band and the CN violet system and implemented into MassiveOES and available as Supplementary Data are described in sections \ref{Swandata} and \ref{sec:CN}.

\subsection{Thermal population distributions}
\label{sec:thermMB}

The distribution of energy between the internal degree of freedom of a diatomic molecule is described by the following partition function \cite{barklem2016partition}

\begin{multline}
\label{Qpart}
Q=\sum_e\sum_{\nu=0}^{\nu_{max}}\sum_{J=\Lambda}^{J_{max}}g_{\Lambda,hfs}\left(2S+1\right)\left(2J+1\right)\\ \times \exp\left[-\frac{hc}{k_BT}\left(T_e+G(\nu)+F_\nu(J)-E_0\right)\right]
\end{multline}

where $e$ denotes electronic states with projection of orbital angular momentum quantum number $\Lambda$ and total spin quantum number $S$, and $\nu$ and $J$ are the vibrational and rotational quantum numbers, respectively. $g_{\Lambda, hfs}$ is the statistical weight factor related to $\Lambda$-doubling and the nuclear hyperfine structure. $T_e$ , $G(\nu)$ and $F_\nu(J)$ represent the electronic, vibrational and rotational energies in wavenumber units. $E_0$ is the energy of the lowest state of the molecule (i.e. $\nu=0$, $J=\Lambda$ for
the ground electronic state). $T$ is the temperature of the heat's bath in the case where all internal degrees of freedom of the molecular gas are thermalized.

Assuming that the internal degrees of freedom are independent, which is already implicitly assumed in the formalism used to derive equation \ref{Qpart},  \footnote{While defining the energy contributions of all internal degrees of freedom to a state defined by the quantum number $\nu, S, J$ and $\Lambda$ as additive, we make the assumption that those degree or freedom are independent \cite{mcquarrie1973statistical}.} the partition function for rotational and vibrational states can be rewritten as

\begin{equation}
\label{QnuJ}
Q(\nu, J)\cong Q_{rot}(J, T_{rot})Q_{vib}(\nu, T_{vib})
\end{equation}

where $T_{rot}$ and $T_{vib}$ are the temperature for the rotational and vibrational distribution functions in the case that they do not equilibrate with each other but obey Boltzmann distribution functions.

Using the Maxwell-Boltzmann distribution law, stating that the number of molecules $dN_E$ having an energy between $E$ and $E+dE$ is proportional to $e^{(-E/k_BT)}dE$ and using equations \ref{Qpart} and \ref{QnuJ}, the population density for rotational and vibrational levels can be defined as \cite{herzberg1950molecular, mcquarrie1973statistical}

\begin{equation}
\label{NJMB}
N(\nu, J)=\frac{N(\nu)}{Q_{rot}}\left(2J+1\right)\exp\left(-\frac{F_\nu(J)}{k_BT_{rot}}\right)
\end{equation}

\begin{equation}
\label{NvMB}
N(\nu)=\frac{N_e}{Q_{vib}}\exp\left(-\frac{G(\nu)hc}{k_BT_{vib}}\right)
\end{equation}

where $N_e$ is the total density of the electronic state $e$ and $N(\nu)$ is the density of the $\nu$ level summed up over its rotational distribution.

Assuming a Boltzmann distribution for the rotational and vibrational levels, the intensity of a line for a transition $(\nu' \rightarrow \nu'' , J' \rightarrow J'',\Omega' \rightarrow \Omega'')$ is given by 
\begin{multline}
\label{eq:Ilambda}
I(\lambda) =  \frac{N_{e}}{Q} e^{-\frac{E(\nu')}{k_b T_{vib}}} (2J'+1) e^{-\frac{E(J',\Omega')}{k_b T_{rot}}} A_{\nu' \nu'' J' J' \Omega' \Omega'''}\cdot\psi(\lambda-\lambda_0)
\end{multline}

where $\psi(\lambda-\lambda_0)$ is the convoluted line profile and instrumental profile of the spectrometer and $\lambda_0$ the central wavelength position of the line. $E(\nu)=G(\nu)hc$ and $E (J',\Omega')\approx F_\nu(J')$ where the effect on energy level splitting due to $\Lambda$-doubling is expressed as a function of $\Omega$ (cf. equation \ref{eq:Ilambda}). We use here the convention following Herzberg \cite{herzberg1950molecular} that the rotational and vibrational quantum numbers of the upper level are defined by the index J', $v'$ and the lower level is described by J'' and $v''$.

\subsection{Swan band database: description, implementation, benchmark}
\label{Swandata}

C$_2$ is a homonuclear diatomic molecule and the Swan band is an electronically allowed dipole transition between two $^3\Pi$ states due to change of symmetry of the wavefunction between the upper and lower state $g\rightarrow u$. These are triplet states and for each rotational quantum number $J$, there are sub-bands $^3\Pi_0-^3\Pi_0$, $^3\Pi_1-^3\Pi_1$ and $^3\Pi_2-^3\Pi_2$. The sub-bands are dominated by P and R-branches (corresponding to selection rule $\Delta J= J' - J'' =-1$ and +1 respectively). Only weak Q-branches ($\Delta J=0$) exist and none for the $^3\Pi_0 - ^3\Pi_0$ sub-band as only $\Delta J=\pm 1$ transitions occur for $\Omega=0$ where $\Omega$ is the total electronic angular momentum along the internuclear axis is designated with $\Omega=\Lambda+\Sigma$. $\Sigma$ is the projection of the spin quantum number S onto the internuclear axis. $\Lambda$ is the quantum number for the component of the total angular momentum along the internuclear axis. It is the absolute value of the projected component of the angular momentum L on the internuclear axis $M_L$ with $\Lambda=\mid M_L\mid$ \cite{SvanbergBook}. $\Pi$ states have by definition $\Lambda=1$ and for the Swan band we have triplet states $d^3\Pi_g$-$a^3\Pi_u$ for which S=1 and so $ \Sigma = M_S =-1, 0, +1$. The coupling between the rotation of the nuclei and $L$ additionally leads to a splitting into two components for each J value when $\Lambda\neq 0$. This is called $\Lambda$-doubling (see below). As the C$_2$ molecule is symmetric and the carbon atom has a zero nuclear spin, transition between symmetric and asymmetric wavefunctions are forbidden. The statistical weight due to nuclear spin I of the anti-symmetric rotational levels is given by I(I+1) and as for C$_2$ we have I=0, they are absent. However, for $\Lambda=1$, we have symmetric and anti-symmetric combination of wave functions due to $\Lambda$-doubling and therefore both even and odd rotational numbers are present. In the P, Q and R branches, only one of the $\Lambda$-doubled levels are present alternatively which leads to the appearance that the even levels are displaced to one side and the odd levels to the other side of a mean position for a given branch progression. With each alternate line that is simply absent because of I=0, this is often referred to as the \lq\lq staggering\rq\rq effect. We note that the statistic weight ($g_{\Lambda,hfs}$ in equation \ref{Qpart}) due to the nuclear spin for individual rotational levels induce an alternance of intensities for sub-bands of electronic transitions with $\Lambda\neq 0$. For more details on the rovibrational structure and lines positions of the $C_2$ Swan band system we refer to Herzberg \cite{herzberg1950molecular} and Pellerin \cite{Pellerin1996}.

\begin{figure}[ht]
\centering
\includegraphics[width=1\textwidth]{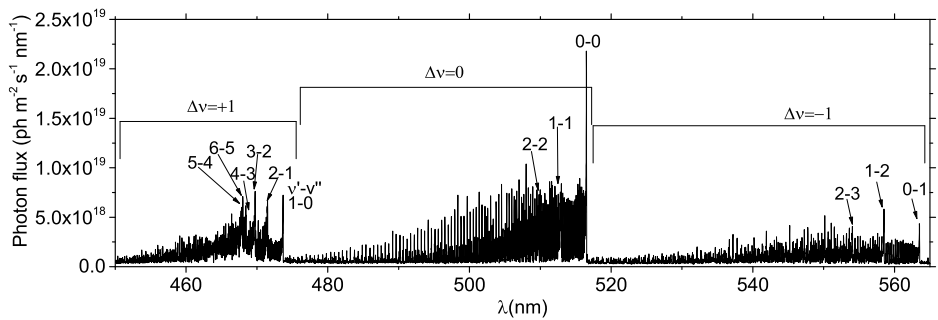}
\caption{Experimental C$_2$ Swan bands emission between 450~nm and 567~nm. The spectrum was acquired in the center of the resonator for a power of 2.7~kW and a flow of 10~L/min.}
\label{fig:C2overview} 
\end{figure}

In figure \ref{fig:C2overview}, a high resolution spectrum from 450~nm up to 565~nm is presented where the main transition groups $\Delta\nu$ and the position of the vibrational transition band heads are identified. Transitions between two electronic states can be ordered following the difference in vibrational quantum numbers during the radiative transition between the upper and lower state $\Delta\nu = v'-v''$. For instance, in figure \ref{fig:C2overview} the transition group $\Delta\nu = +1$ corresponds to the rovibrational serie d$^3\Pi_g (v')$-a$^3\Pi_u (v'-1)$. Note that the d$^3\Pi_g (v'=0)$ level is therefore not part of this transition group. The $\Delta\nu=-2, +2$ transition groups can also be seen experimentally but are not shown here.

The spectroscopic database that we implemented into the fitting program of molecular emission spectra MassiveOES \cite{Vorac2017} is derived from the work of Brooke et al. \cite{BROOKE201311}. As initial input, MassiveOES requires rovibrationally resolved Einstein coefficients for the transitions between the upper and lower state that can be generally calculated using the following equation \cite{Bernath_book}:

\begin{equation}
\label{Acoeff}
A_{(v', J')\rightarrow (v'',J'')}=\frac{16\pi\nu^3S_{J''}^{\Delta J}}{3\epsilon_0hc^3\left(2J'+1\right)}<\Psi_{v'J'}\mid R_e(r)\mid\Psi_{v''J''}>^2
\end{equation}

where $S_{J''}^{\Delta J}$ is the H\"{o}ln-London factor and $<\Psi_{v'J'}\mid R_e(r)\mid\Psi_{v''J''}>$ is the transition dipole moment (TDM).

Brooke et al. \cite{BROOKE201311} used PGOPHER, a program developed by Colin Western, for calculating the Einstein $A_{(v', J')\rightarrow (v'',J'')}$ coefficients \cite{Western2017pgopher}. PGOPHER requires as input a set of molecular constants and the band strengths for each vibrational band $v'\rightarrow v''$.  The purely vibrational transition dipole moments (TDMs) were calculated using the program LEVEL \cite{LEROY2017167}, a program that can solve the 1D Schr\"odinger equation for diatomic molecules when provided with a potential energy curve and the electronic TDM. The set of molecular constants for fitting RKR potential curves of the upper and lower electronic levels involved into the Swan band transition are listed in \cite{BROOKE201311}. With the output of LEVEL, PGOPHER is then able to calculate the rotational TDMs and the H\"{o}ln-London factors required to compute the numerical values of equation \ref{Acoeff}. The original simulations by Brooke et al were extended to a maximum rotational quantum number J of 300 by running again PGOPHER with the same molecular constants. This allows to properly calculate the rotational partition function in MassiveOES for the high rotational temperatures encountered in the present study. It should however be noted that, while doing this, the calculated energy levels have increasing error for increasing J in energy and line positions. The molecular constants used for the C$_2$ Swan band are given in Appendix \ref{C2appendix} and the output calculation of PGOPHER as Supplementary Material to this paper.

\begin{figure}[ht]
\centering
\includegraphics[width=0.7\textwidth]{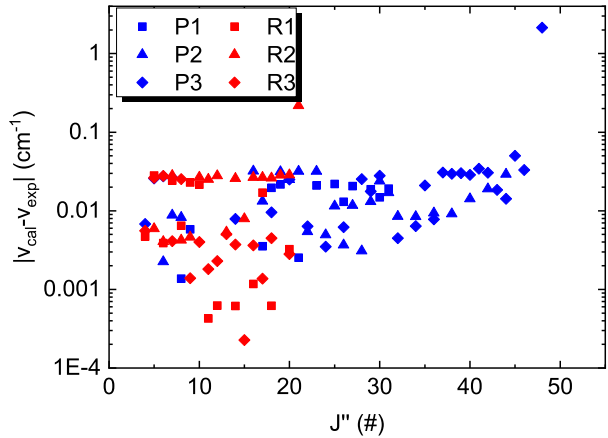}
\caption{Difference between the calculated line position $\nu_{cal}$ and the experimental one $\nu_{exp}$ as function of the rotational quantum number of the lower state. Red and blue dots indicates the P and R branches respectively. Square, circle and diamond indicate the different fine structure sub-bands. All lines comes from the $\nu'=0$ level and only the main transitions are considered which corresponds to $\Delta\Omega=0$.}
\label{fig:linepositiontest}
\end{figure}

Extensive comparison of calculated line positions versus experimental ones were already done in the literature (see Brooke et al. \cite{BROOKE201311} and references therein). We additionally performed a cross-check with the high resolution degenerate four-wave mixing experiments of Lloyd and Stewart \cite{Lloyd1999}. The difference between their measured line positions for the $\Delta \nu=0$ transition group and our calculated ones are shown in figure \ref{fig:linepositiontest}. One can see that, excepted for two outliers, the line positions are calculated with an accuracy much better than $\pm$ 0.05 cm$^{-1}$. Experimental data for higher J numbers is however not available for performing further benchmark. Experimentally we have a line profile of 22~pm corresponding to about 0.8 cm$^{-1}$ which is much broader than the line position uncertainty $\lambda_{err} <1$ pm. Consequently, using calculated line positions for the fitting of the experimental spectra induces no additional error related to uncertainties in line position assignment. We note at this point that high resolution measurements of line positions were done up to $J=80$ and only for some vibrational levels by Tanabashi et al. \cite{Tanabashi2007}. 

The database of lines and their transition probabilities is compiled in a SQL database, that is then read by MassiveOES. The database consists of three tables for listing the lines, the identity of rovibrational \lq\lq upper\rq\rq\, and \lq\lq lower\rq\rq\, states respectively. Each element of the table corresponds to a single line in the spectrum for which energy of the transition, air wavelength, vacuum wavelength, Einstein coefficient, upper state identity and lower state identity are given. The \lq\lq upper state\rq\rq~tag (as well as the \lq\lq lower state\rq\rq~tag) links the line with the energy level from which is originated, that will be used to calculate the intensity of the line. The effect of molecular perturbations on the energy level and transition probability of the $d^3 \Pi_g$  $\nu$= 4 and 6  is included in the Einstein coefficient and energy of the transitions. The extra lines due to the perturbations for the  $d^3 \Pi_g$ state by the $b^3\Sigma_g^- (\nu=16,19)$ and $1^5\Pi_g$ states \cite{BROOKE201311} are combined in the table as additional transitions from the upper state, such that, the rovibrational population distribution is calculated without extra energy levels.

\subsection{CN violet system emission}
\label{sec:CN}


The CN molecular violet system corresponds to the $B^2\Sigma^+ \rightarrow X^2\Sigma^+$ transition of CN which is an heteronuclear diatomic molecule. The required molecular constants have been taken from S. Ram et al. \cite{RAM2006225}. The transition probabilities between two vibrational states calculated by solving the integral of the vibrational wavefunctions of the upper and lower states and TDM $\int \Psi_\nu(r) R_e(r) \Psi_{\nu'}(r) $ were extracted from LIFBASE \cite{luque1999lifbase}. The  molecular constants for calculating the Rydberg-Klein-Rees (RKR) potential curves of the X$^2\Sigma^+$ and B$^2\Sigma^+$ states were taken from \cite{CERNY1978154, ITO1988}). For the TDMs $R_e(r)$, the values of Ito et al. \cite{ItoCN1991} were taken. These molecular constants were then used as input for PGOPHER to calculate the H\"{o}ln-London factors of the CN violet system with J number up to 180. For the CN violet system some perturbations also exist but those are not included in the calculations as they do not affect significantly the spectra. In Appendix \ref{CNappendix} are given all the constants used in the calculation. As sanity check, a one to one comparison between our PGOPHER calculations and LIFBASE was made and they yield the same results both for line positions and intensities. 

\section{Spectral analysis of the C$_2$ Swan band}

\subsection{Non-Boltzmann fitting of the Swan band} 
\label{StatebState}


In non-equilibrium plasmas, rotational and vibrational distributions do not always thermalize because of competition between production and destruction processes and a molecular spectrum can then not be fitted using single rotational $T_{rot}$ and vibrational T$_{vib}$ temperatures \cite{Bruggeman2014RotT}. As discussed in section \ref{C2kin}, the C$_2$ ($^3\Pi_g$) state is well known for overpopulations in the density of its high vibrational levels and particularly the $v=6$ level. In this section we investigate high resolution spectra of the C$_2$ Swan band in order to assess possible deviations from thermal equilibrium of the rovibrational populations. 


\begin{figure}[ht]
\centering
\subfloat[][Identification of individual lines from the P and R branches of the v=0 level.]{\includegraphics[width=0.95\textwidth]{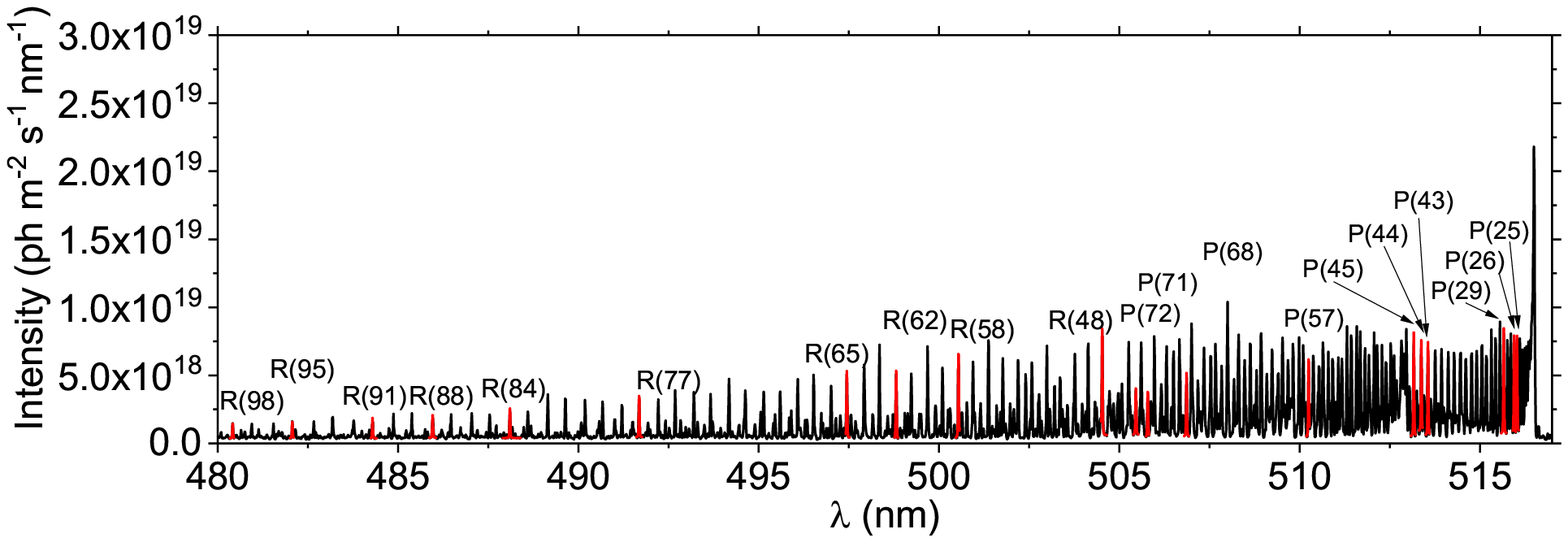}}
	
\subfloat[][Boltzmann plot constructed using selected lines of the $\Delta\nu=0$ transition group.]{\includegraphics[width=0.7\textwidth]{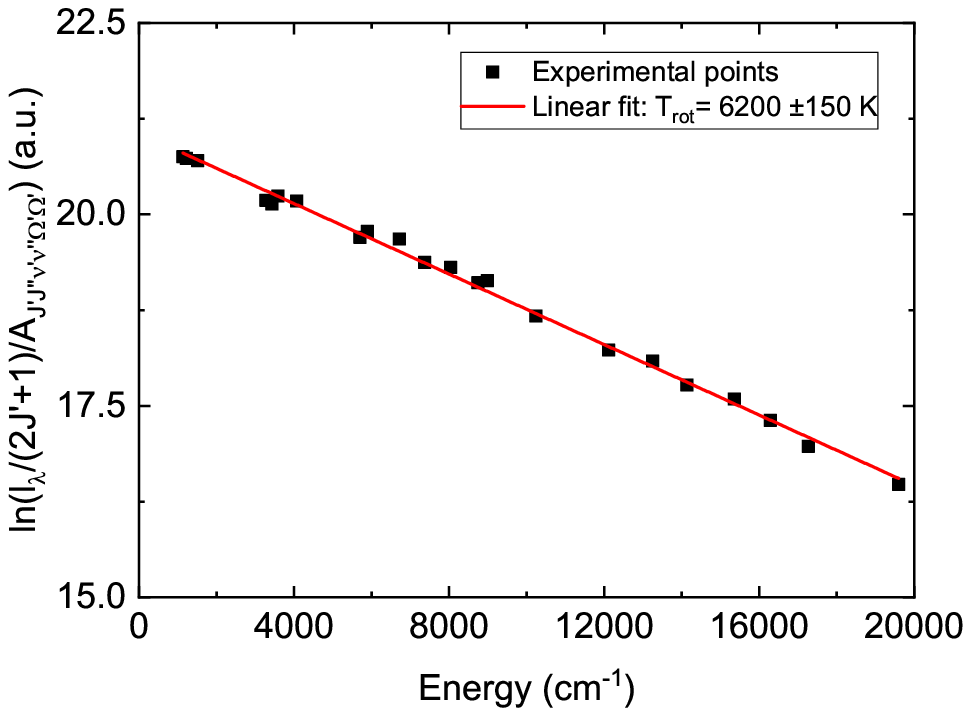}}
\caption{Selection of isolated lines of the $\Delta \nu=0$ C$_2$ Swan band and Boltzmann plot fitting to obtain the rotational temperature of the $d^3\Pi_g$ state. Experimental conditions are: 2.7 kW input power, 10 L/min of CO$_2$, an operating pressure of 920 mbar. The measurements are performed at a height of 20~mm (i.e. middle) inside the resonator and for the radial position r=0~mm.}\label{fig:linebyline}
\end{figure}

%

In figure \ref{fig:linebyline}, individual P and R lines from the $v=0$ level are identified (the Q lines are too weak to be distinguished, see section \ref{Swandata}) from the $\Delta\nu=0$ transition group and their area fitted with a Voigt profile (with gaussian and lorentzian half-width of respectively $\sigma_G$=9.5 pm and $\sigma_L$= 4 pm) to take into account the apparatus profile determined from Hg lines using a low pressure Hg lamp. They were selected only when they can be separated from neighboring lines and clearly identified. Note that the fine structure of the rotational levels could not be resolved and so that the lines are weighted over the rotational fine structure (i.e. rotational sub-bands). The logarithm of the density of each individual rotational level is plotted as function of the rotational energy (i.e. Boltzmann plot). All levels fit on a single line which can be fitted with a rotational temperature of 6200$\pm 160$~K. Here and in the following, we will quote for each fitting method, its related statistical errors. No deviation is observed neither for low or high rotational levels indicating that the levels are thermalized. We note here that levels with approximately J$>$120 do not contribute noticably  to the emission spectra and are not visible.


\begin{figure}[ht]
\centering
\subfloat[][\label{fig:nu0fit_residues}]{\includegraphics[width=0.55\textwidth]{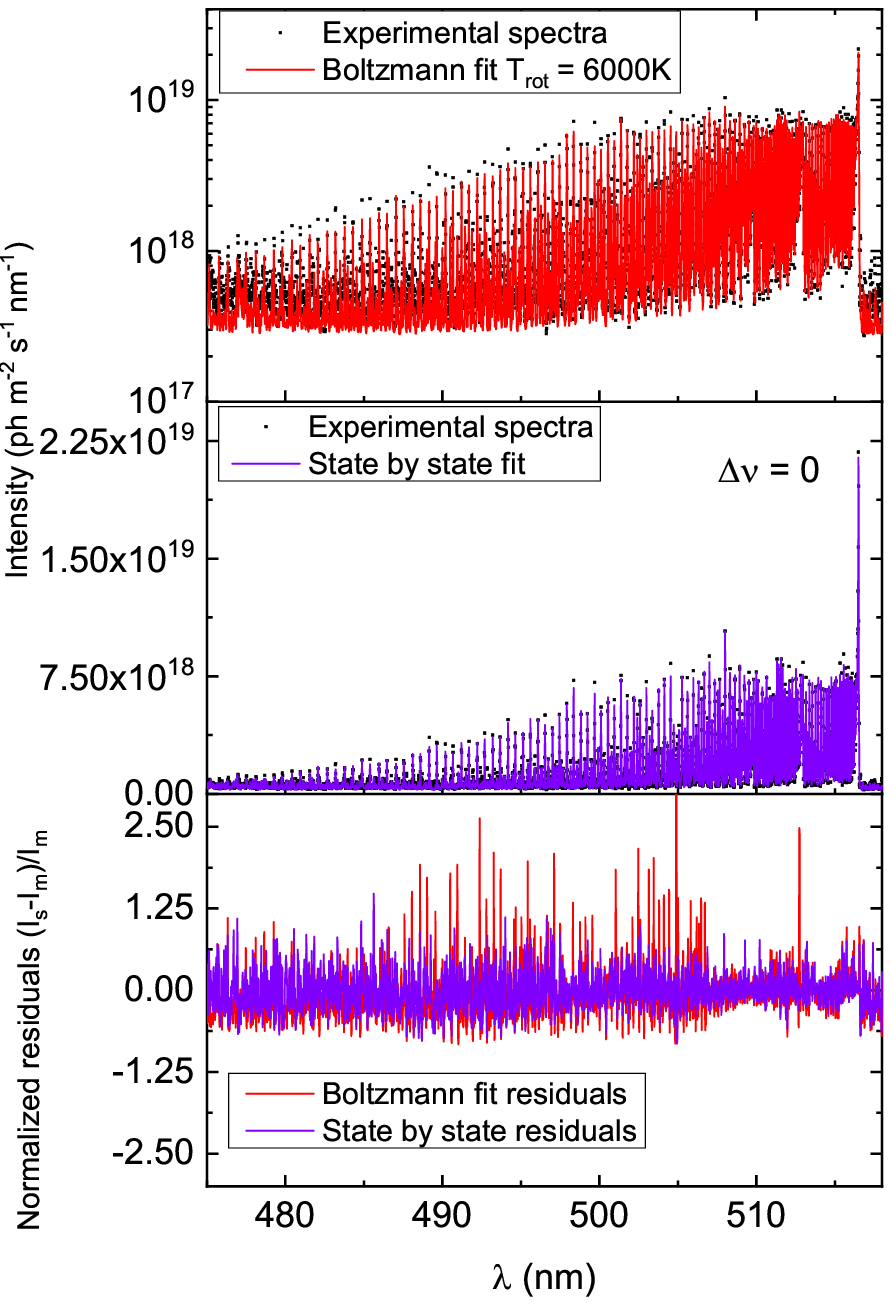}}
\subfloat[][\label{fig:statebstateNu0boltz}]{\includegraphics[width=0.55\textwidth]{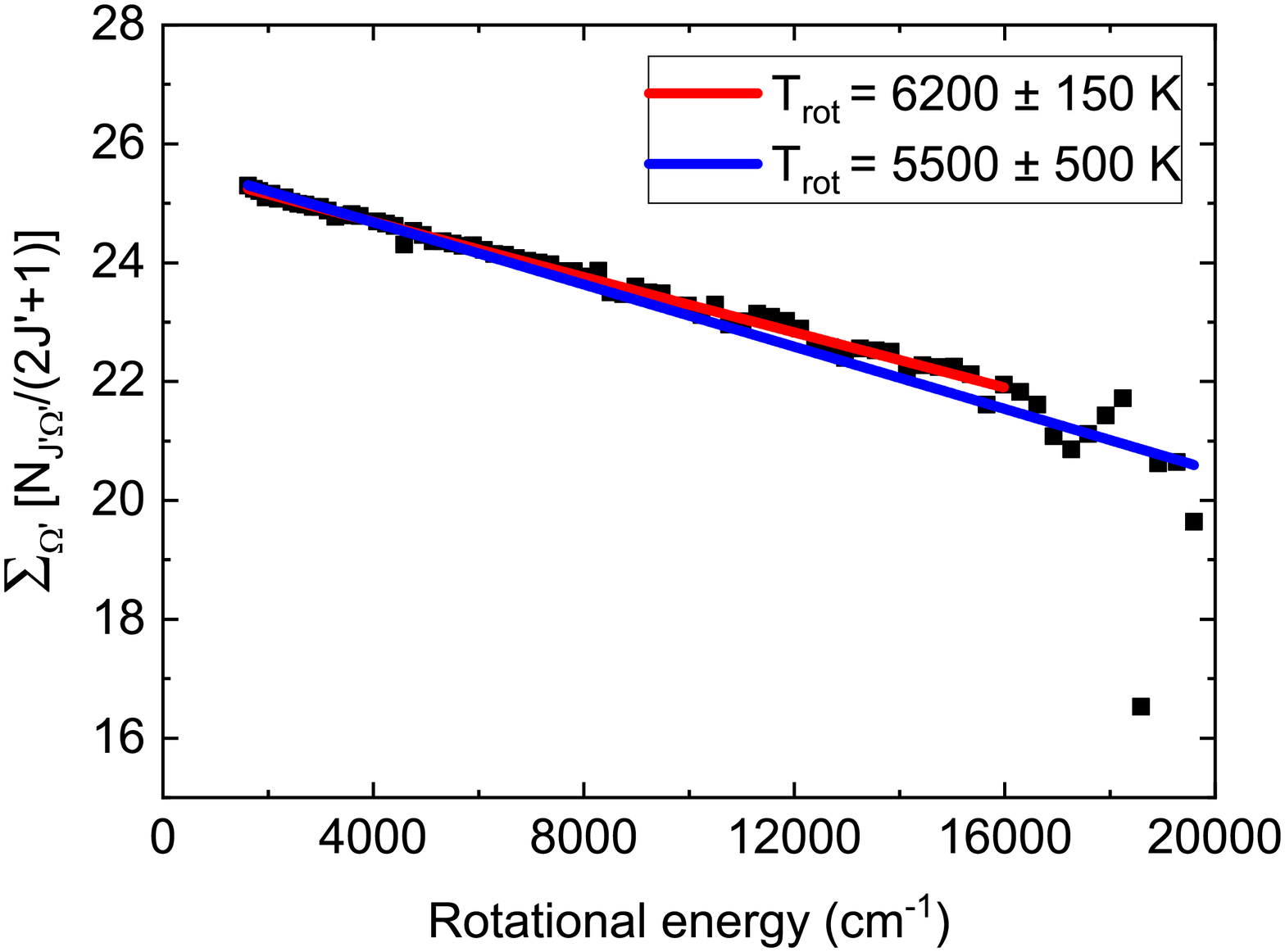}}
\caption{Analysis of the $\Delta \nu=0$ transition group of the C$_2$ Swan band: a) Comparison between Boltzmann and state by state fitting. Note the logarithmic and linear scales used for presenting the fitted spectra. b) Boltzmann plot of the rotational population densities obtained from the state by state fitting. The experimental conditions are the same as in figure \ref{fig:linebyline}. $I_m$ and $I_s$ stand respectively for the intensity of the measured and simulated spectra.}
\label{fig:statebstateNu0}
\end{figure}

The thermalization of the rotational levels of the $d^3\Pi_g$ state is further assessed using the option in MassiveOES of state-by-state fitting \cite{Vorac2017b}. State by state fitting here means that the fitting routine evaluates the density of each upper rotational state while minimizing the residue for all transitions from the same upper state (i.e. different transition groups, P, Q and R branches and $\Lambda$-doubling). Allowing to fit independently the density of all rotational states helps to reduce the overall residue of the fit but only marginally. In figure \ref{fig:statebstateNu0}, the population density of the rotational levels versus their energy is plotted. All rotational population densities fit quite well on a single line indicating no deviation from a thermal rotational distribution function. The lowest rotational levels are not shown because they are experimentally too much overlapped and the state by state fitting routine cannot converge (see below for a discussion of those). The highest rotational levels are widely scattered due to the very low signal to noise ratio. A rotational temperature T$_{rot}$=5500$\pm$ 500~K is obtained from the fitting of figure \ref{fig:statebstateNu0}. From this first fit, some systematic deviations are apparent. A closer inspection indicates the presence of some outliers points which have densities off by an order of magnitude or more without any systematic trend. Removing these outliers which have a non-neglible weight for the fitting of a linear slope in logarithmic plot, the statistical uncertainty of the Boltzmann plot is significantly reduced. The inverse slope of the Boltzmann plot gives a temperature of 6200$\pm$150~K. This result agrees with the line by line Boltzmann plot analysis made in figure \ref{fig:linebyline}. Fitting the complete $\Delta\nu=0$ transition group while assuming a Boltzmann distribution (i.e envelope fitting) for rotational and vibrational distribution gives similarly a temperature of 6000$\pm$200~K in excellent agreement with the state to state fitting analysis. The result of the fit using a Boltzmann distribution is also shown in figure \ref{fig:statebstateNu0} (note the logscale) as well and one can see that the residues following a Boltzmann fitting or a state by state fitting are similar.

\begin{figure}[ht]
\centering
\subfloat[][Rotational population densities for \newline \hspace*{1.5em} the $\nu=0$, 1, 2 vibrational levels.\label{fig:statebstateAll}]{\includegraphics[width=0.55\textwidth]{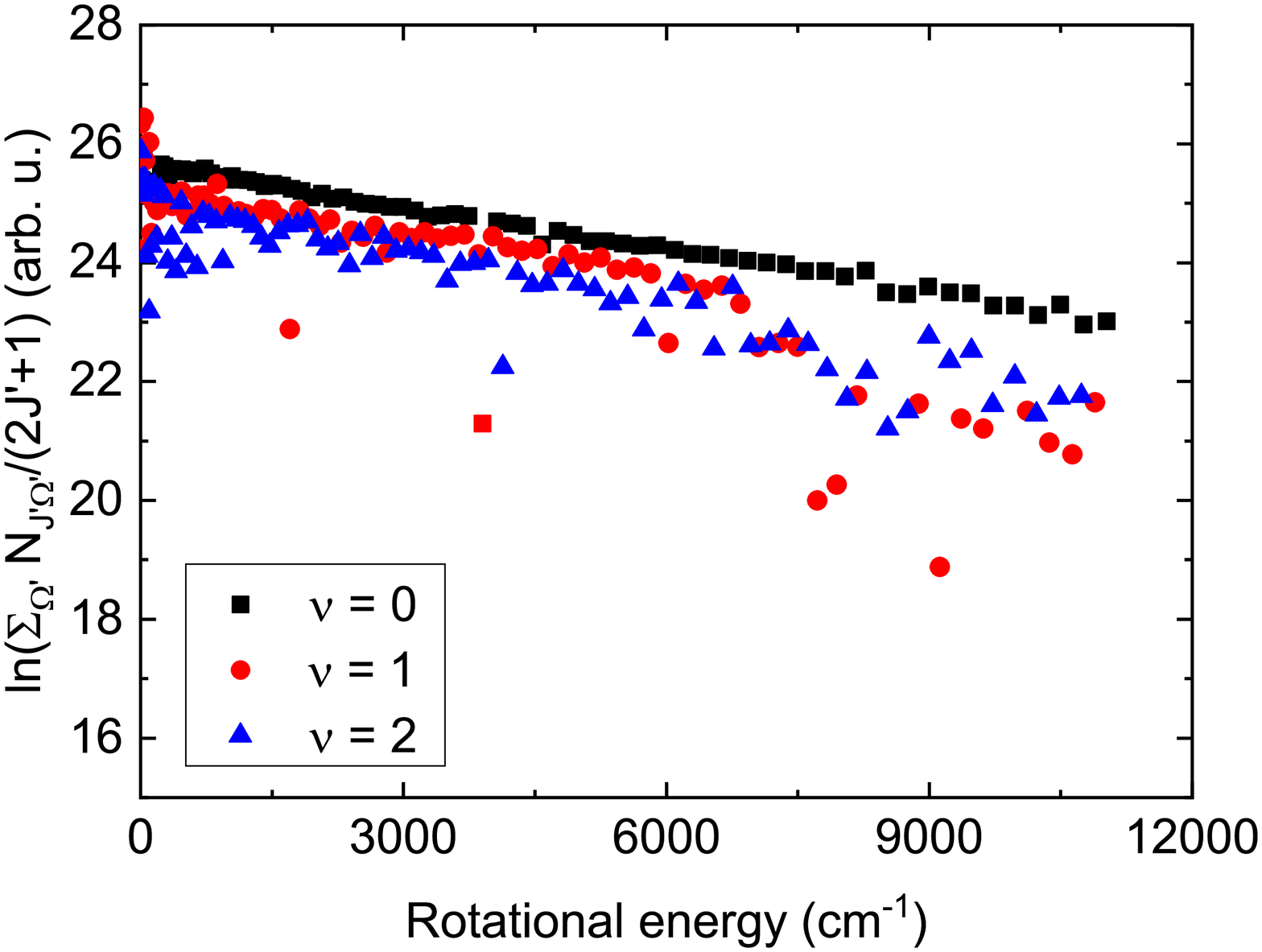}}
\subfloat[][Vibrational distribution function of \newline \hspace*{1.5em} the C$_2$ $d^3\Pi_g$ state.\label{fig:VibStatebyState}]{\includegraphics[width=0.55\textwidth]{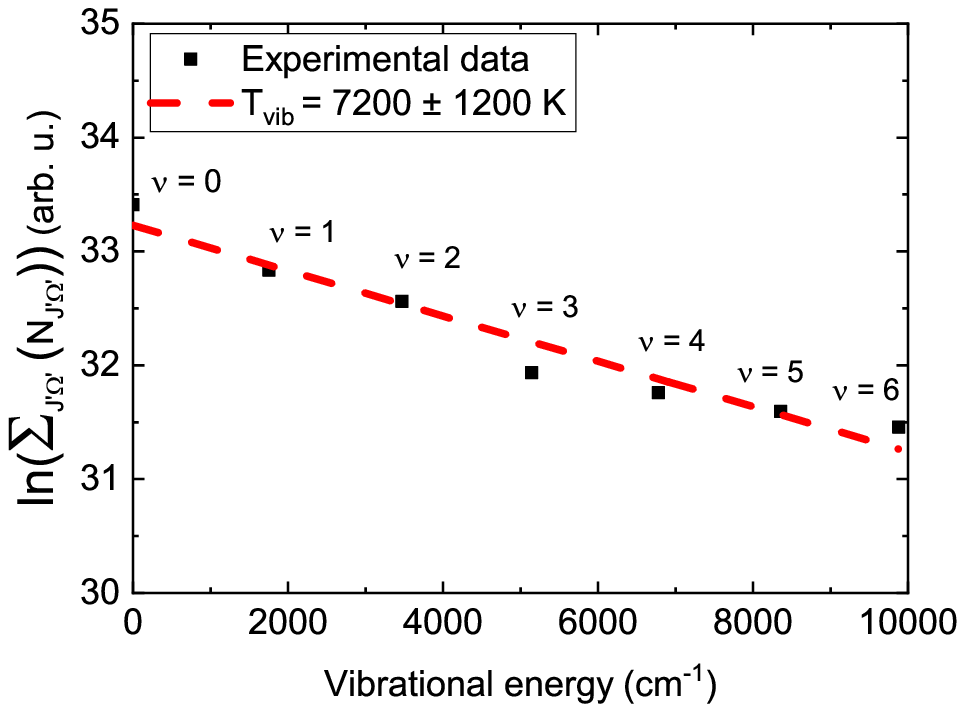}}
\caption{Rovibrational densities of the $d^3\Pi_g$ state obtained from the state by state fitting analysis of the $\Delta \nu=-1, 0$ and $+1$ transition groups of the C$_2$ Swan band. The experimental conditions are the same as in figure \ref{fig:linebyline}.}
\end{figure}

In figure \ref{fig:statebstateAll}, the state by state fitting analysis is extended to the $\Delta\nu=-1$ and $\Delta\nu=+1$ transition groups for the rotational and vibrational states of the $d^3\Pi_g$ state. One can see that the rotational levels of the v=1 and v=2 levels also follow a Boltzmann distribution with the same slope as for the $\nu=0$ level although the signal is more noisy. Also population densities of the lower rotational states can be analysis and no systematic deviations from a single slope can be found.

\begin{figure}[ht]
\centering
\includegraphics[width=0.7\textwidth]{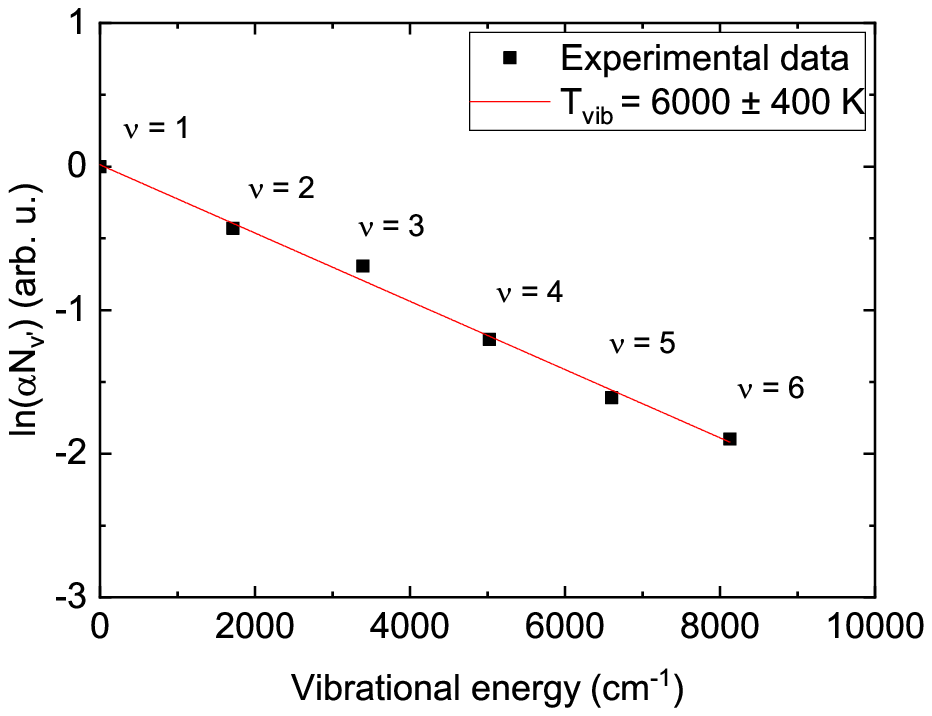}
\caption{State by state fitting of the vibrational distribution function of the C$_2$ $(d^3\Pi_g)$ state using the $\Delta\nu=+1$ band. The rotational temperature was fixed to 6200~K and the experimental spectrum constructed by adjusting the density of the $v$ levels. The experimental conditions are the same as in figure \ref{fig:linebyline}.}
\label{fig:VibStatebyStateManual} 
\end{figure}

After integration over the rotational distribution the density of the vibrational states can be obtained (cf. equation \ref{NJMB}). The resulting Boltzmann plot for the vibrational states is shown in figure \ref{fig:VibStatebyState}. Their population distribution can be described by a single temperature of 7200$\pm$ 1200~K. For comparison, we also performed a state by state fitting of the $\Delta\nu=+1$ transition group that is the most sensitive to the vibrational temperature (see section \ref{Sec:sensitivity}). The rotational temperature was fixed to 6200~K following the analysis of the $\Delta\nu=0$ transition group. For the $\Delta\nu=+1$ transition group, the density of each $\nu$ level is adjusted independently and the results are shown in figure \ref{fig:VibStatebyStateManual}. One can see that all vibrational levels fit perfectly on a single line with T$_{vib}$=6000$\pm$ 400~K. The larger uncertainties in the case of the state by state fitting when all three transition groups $\Delta\nu=+1, 0, -1$ are taken into account stems from the lower signal to noise ratio for the $\Delta\nu=-1$ transition group. The result of the fitting of the 3 transition groups for the vibrational distribution of the C$_2$ d$^3\Pi_g$ state shown in figure \ref{fig:VibStatebyState} may let someone thing that the C$_2$ Swan band system follows a two-temperature vibrational distribution function. The vibrational state to state analysis of the most sensitive $\Delta \nu = +1$ transition group done in figure \ref{fig:VibStatebyStateManual} (which is more sensitive to T$_{vib}$ as discussed in section \ref{Sec:sensitivity}) show that there is no evidence of a two-temperatures vibrational states distribution function. Fitting the $\Delta \nu = +1$ transition group alone, which is the one usually measured in the literature for studying the so-called C$_2$ high pressure bands as it is the most sensitive for determining vibrational states densities, show that the vibrational distribution function of the d$^3\Pi_g$ state is very well described by a single temperature. Considering the noise in the state by state ro-vibrational fitting shown in figure \ref{fig:statebstateAll} for $\nu=1$ and 2 levels, it is in fact quite remarkable to see that the fitted vibrational temperature still agrees with the more precise fitting given in figure \ref{fig:VibStatebyStateManual} within their own error bars. Note that for the vibrational state to state fitting of the $\Delta \nu = +1$ transition group, the rotational temperature was set using the T$_{rot}$ obtained from the more sensitive $\Delta \nu = 0$ transition group.


From this analysis, it appears that for the present plasma conditions, the vibrational levels are completely thermalized and no overpopulation of the higher vibrational levels and more particularly v=6, occurs. Analysis of other spectra as the one discussed above yield the same results and no deviations from Boltzmann equilibrium either for the rotational or vibrational levels could be found for the conditions of the present study (cf. section \ref{sec:results}). Either the processes discussion in section \ref{C2kin} leading to the so-called C$_2$ high pressure bands are absent in the present discharge or the levels thermalize within the effective (radiative) lifetime of the state (see section \ref{C2relax} for a discussion). Different methods of fitting give similar results and no evidence of deviations from thermal distributions can be found within error margins. For the following, Boltzmann distributions are then assumed and an envelope fitting of the spectra is done. For measuring the rotational temperature, the $\Delta\nu=0$ transition group is chosen while all three transition groups are used for the determination of the vibrational temperature. For more discussion on this, we refer to section \ref{Sec:sensitivity}.

\subsection{T$_{rot}$ and T$_{vib}$ from the C$_2$ Swan band: sensitivity and uncertainty analysis}
\label{Sec:sensitivity}

The convergence of a fit is usually assessed by looking at the value of the least mean square error and minimizing the root mean squared error ($RMSE=\sqrt{\sum_i (y_i-\overline{y_i})^2}$). It is possible to use the computed residue for assessing how sensitive is the minimization routine for given external parameters that are being fitted. Information can be extracted from a dataset (here an experimental spectrum) only when the variation of signal is statistically significant. It usually means that the correlation between a signal and the event we want to measure must be significantly larger that the statistical noise. Theoretical spectra can be computed for given fixed rotational and vibrational temperatures and compared with another theoretical spectrum in terms of their respective RMSEs \cite{Staack2006}. In the case where the (thermal) white noise is the dominant source of uncertainty in the fitting of an experimental spectrum, the amplitude of the noise can be directly correlated to an iso-line inside which the \lq\lq true value\rq\rq~ stands. Nassar \cite{Nassar2014} evaluated the errors induced by a white noise on T$_{rot}$ determination while fitting the band head of the $\Delta\nu=0$ transition group and obtained errors in the order of the signal to noise ratio or lower. This is however without considering systematic errors that are induced by performing mathematical operations on the data such a baseline correction or intensity calibration. These errors are additive and amplify the uncertainties on the values obtained through a fit. Also, one needs to take additionally into account the uncertainties in the molecular constants (Franck-Condon and H\"{o}ln-London factors).

\begin{figure}[ht]
\centering
\includegraphics[width=0.55\textwidth]{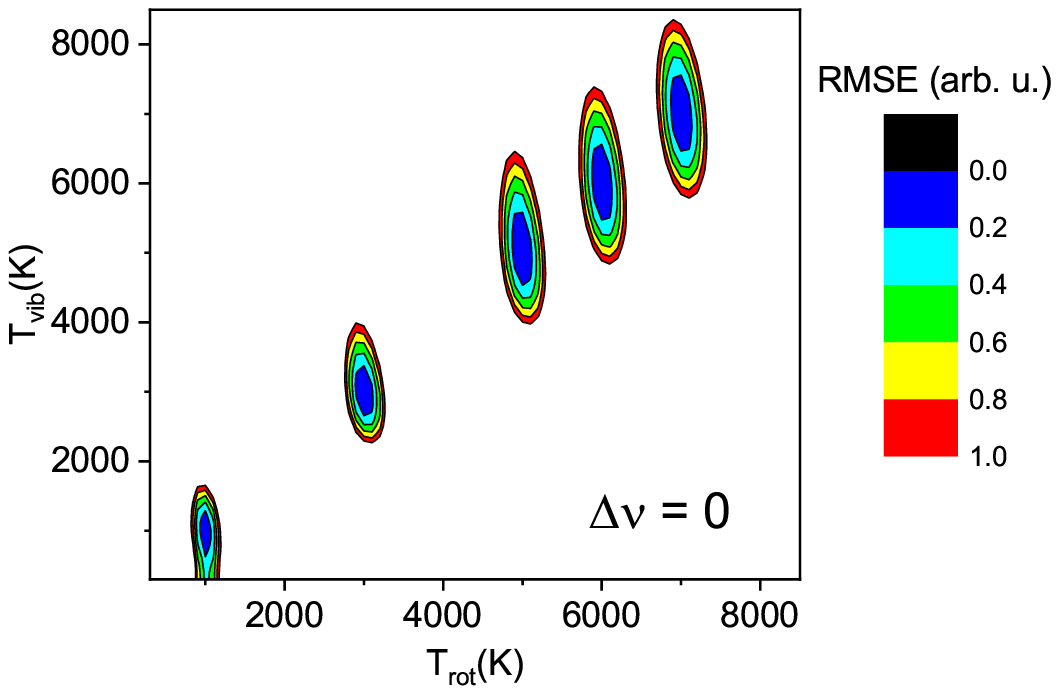}\includegraphics[width=0.55\textwidth]{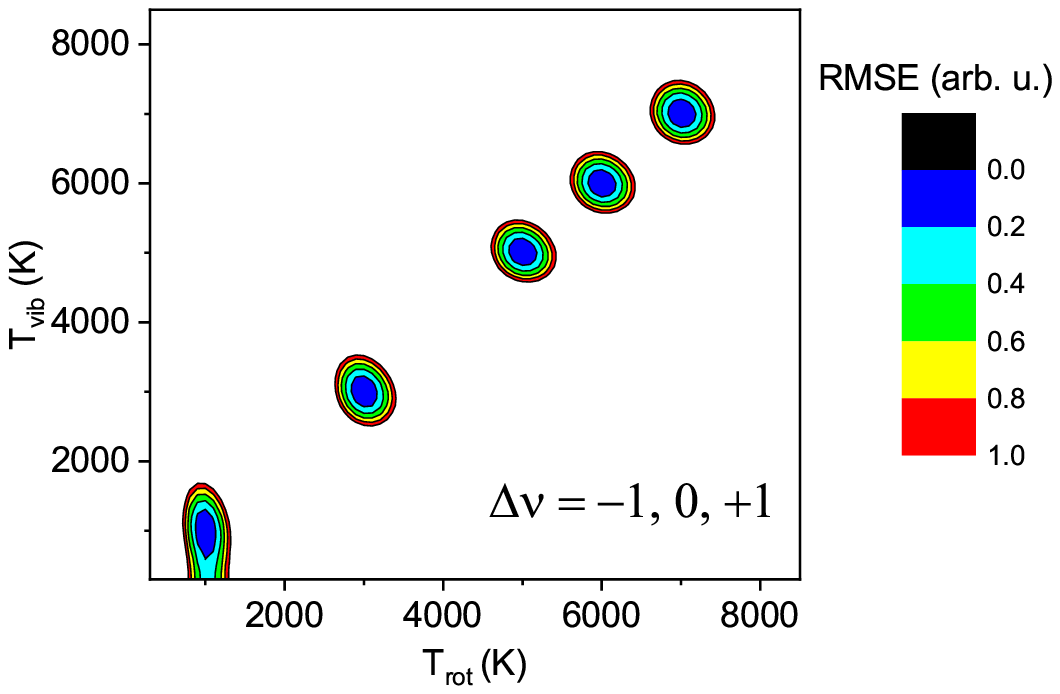}
\caption{Sensitivity of the $\Delta \nu=0$ transition group (475-517 nm) and of the $\Delta \nu=-1, 0, +1$ transition groups (450-567 nm) of the C$_2$ Swan band while fitting simultaneously the rotational and vibrational temperature of the $d^3\Pi_g$ state. The calculations are done against a theoretical spectrum using a Voigt profile of 20~pm (equal gaussian and lorentzian contributions) for T$_{rot}$= T$_{vib}$ fixed at values of 1000, 3000, 5000, 6000 and 7500 K.}
\label{fig:sensitivity_contour_theo} 
\end{figure}

In figure \ref{fig:sensitivity_contour_theo}, such analysis is first made for the $\Delta\nu=0$ alone and for the $\Delta\nu=-1, 0, +1$ transition groups of the C$_2$ Swan band by comparing simulated spectra with theoretical spectra with $T_{rot}=T_{vib}$ taking values of 1000, 3000, 5000, 6000 and 7500~K with 5$\%$ surimposed noise. The steepness of the RMSE around the equilibrium point (i.e. true value) obtained by the minimization routine is a good indication of accuracy with which a parameter can be fitted (here $T_{rot}$ and $T_{vib}$). It appears clearly that the $\Delta\nu=0$ is rather imprecise for the determination of the vibrational temperature of the $d^3\Pi_g$ state and that typical errors are in the order of several 1000~K. The minimum for the rotational temperature is on the contrary quite steep indicating that the band has a good sensitivity for $T_{rot}$ determination. The $\Delta\nu=-1, +1$ transition groups have on the other hand a much higher sensitivity to $T_{vib}$. A comparison between the $\Delta\nu=0$ fit and the fit of all three transition groups seems to indicate at first sight that both allow to obtain the rotational temperature with rather good sensitivity. However, in the case of $\Delta\nu=-1, 0, +1$ transition groups fit, one can see that the isolines are stretched on a diagonal (i.e. not perpendicular to the axes) which means that $T_{rot}$ and $T_{vib}$ are correlated during the minimization (i.e. if an error is made in the parameters minimization used for the determination of one variable, the error will propagate to the other). The parametrization of the spectrum in terms of independent variables ($T_{rot}$, $T_{vib}$, instrumental profile,...) in a feature space (i.e. a space made of independent variables that are combinations of the original variables) is not entirely successful and some of the new variables are not orthogonal. This is a well known issue in the field of Principal Components Analysis (PCA) \cite{shlens2014tutorial}.  In the present case, when the isolines are equidistant in the vertical and horizontal directions, one can say that the determination of $T_{rot}$ and $T_{vib}$ are statistically independent. Note that a fitting routine gives for any given fit a RMSE but its absolute value carries no intrinsic information. It has a meaning only in correlation with other RMSEs calculated around the (local) equilibrium point of the fit. The RMS error is then directly normalized to the equilibrium value.

While fitting an experimental spectrum using the assumption of Maxwell-Boltzmann distribution for the rotational and vibrational states of the radiating molecule, there are additional parameters that need to be fixed. They are listed into Table \ref{tab:parameters}. These parameters, excepted for the first three in the table are all experimental and depends of the spectrometer. 
A linear baseline is fitted with an offset value B$_{base}$ and a slope $\alpha_{base}$ for substracting contributions related to noise and continuum emission from the plasma. The instrumental profile of the spectrometer is taken into account via $\sigma_L$ and $\sigma_G$ while the wavelength calibration of the spectrum is adjusted with a quadratic fit ($\lambda_{start}$, $\Delta\lambda$ and $\Delta\lambda^2$ parameters) and the wavelength $\lambda$ is defined as 

\begin{equation}
\lambda = \lambda_{start} + \Delta\lambda \cdot i + \Delta\lambda^2 \cdot i
\end{equation}  

where $i$ is the pixel number. The density $N$ of the upper excited state has no influence on the evaluation of the temperatures as it is related only to the integrated band intensity. 

The parameters that have an influence on the determination of $T_{rot}$ and $T_{vib}$ are the line profile and the fitting of the baseline. By varying systematically these values and considering uncertainties on the Franck-Condon (FC) and H\"{o}ln-London factors, we estimated conservative uncertainty margins of about $\pm 500$ ~K for the spectra fitted in the present study. This evaluation is confirmed by the analysis made in section \ref{StatebState} where statistical methods have been applied while using different analysis methods to evaluate the rovibrational distribution functions. The errors for the rotational temperature are in fact probably lower.  Fitting of theoretical low resolution spectra generated using SPECAIR \cite{Specair, caillaut2006modeling} or from Fantz and W{\"u}nderlich \cite{fantz2014fundamental} leads to difference in the order of 1\% for the rotational temperature. For the vibrational temperature, we also performed a comparison using the transition dipole moments of Kokkin et al \cite{Kokkin2007} and  FC-factors obtained by Fantz and W{\"u}nderlich \cite{fantz2014fundamental}. The main difference with the present values and those calculations are only in the absolute values of the FC-factors with a difference of about 20$\%$. The latter difference is systematic and consequently only affects the determination of the density $N$ of the upper electronic state (i.e. after summation over rotational and vibrational population densities). No significant difference for the determination of T$_{vib}$ while using the two datasets were found.

\begin{table}
\begin{tabular}{c|c}

\hline
Parameter & meaning \\
\hline
$N$ & Upper state excited specie density \\
$T_{rot}$ & Rotational temperature \\
$T_{vib}$ & Vibrational temperature \\
\hline
B$_{base}$ & Constant baseline offset value versus $\lambda$ \\
$\alpha_{base}$ & baseline slope \\
\hline
$\sigma_{G}$ & Gaussian width of the Voigt profile \\
$\sigma_{L}$ & Lorentzian width of the Voigt profile \\
 \hline
$\lambda_{start}$ & Wavelength of the first point in the spectra \\
$\Delta\lambda$ & Distance between two points in the spectra \\
$\Delta\lambda^2$ & Quadratic correction of the position of a point in the spectra \\
\hline
\end{tabular}  
\caption{Total amount of fitting parameters while making a Boltzmann population fit with the MassiveOES fitting program.}
\label{tab:parameters}
\end{table}

\begin{figure}[ht]
\centering
\includegraphics[width=0.55\textwidth]{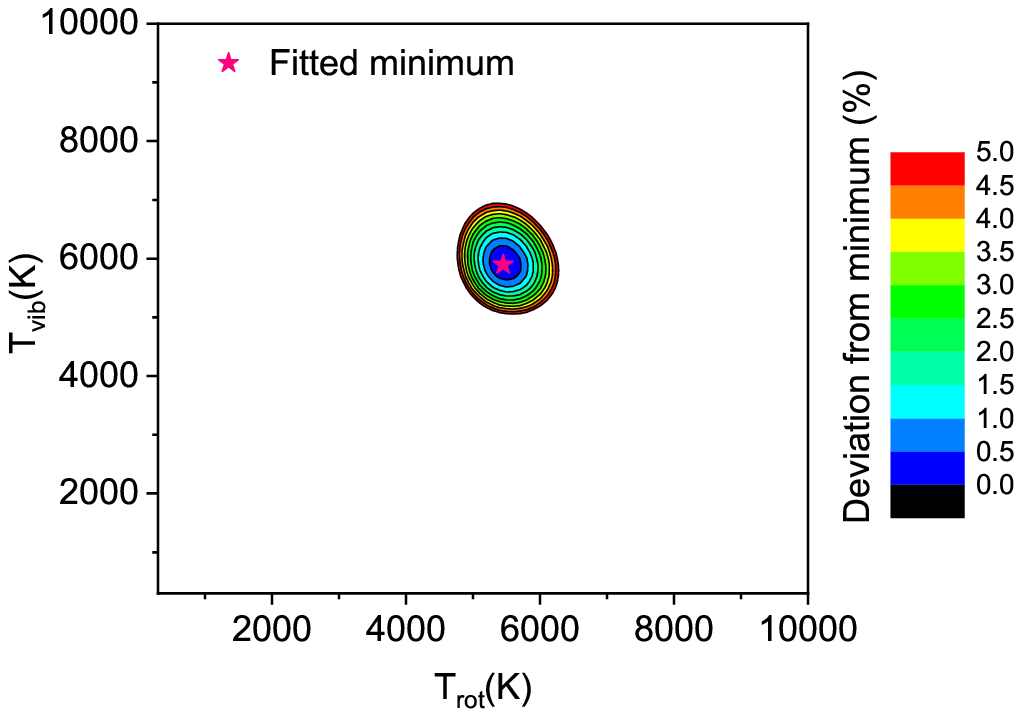}\includegraphics[width=0.55\textwidth]{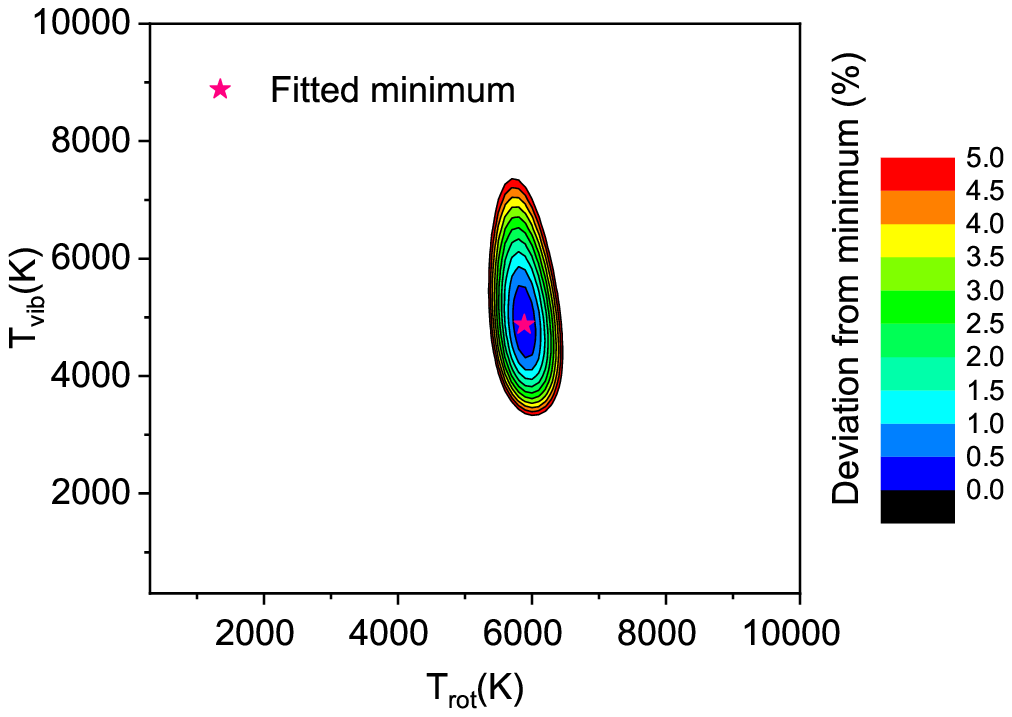}
\caption{Sensitivity of the $\Delta \nu=0$ (on the left)  and $\Delta \nu=0,\pm1$ (on the right) transition groups of the C$_2$ Swan band for fitting simultaneously the rotational and vibrational temperature of the $d^3\Pi_g$ state. The calculations are done for a spectrum acquired with a resolution of about 22~pm.}
\label{fig:sensitivity_contour} 
\end{figure}

In figure \ref{fig:sensitivity_contour}, the same analysis is made for an experimental spectrum. The minimization routine of MassiveOES is here used only for adjusting the area of the spectra (i.e. population density of the electronic state), baseline and linewidth. The spectra are calibrated in wavelength beforehand using the position of known lines and its contribution to fitting uncertainties can be in the present case (i.e high resolution) neglected. The results are similar to figure \ref{fig:sensitivity_contour_theo} and it can be concluded that the C$_2$ Swan $\Delta\nu=0$ transition group is overall more sensitive and \lq\lq precise\rq\rq for the determination of T$_{rot}$. 

\begin{figure}[ht]
\centering

\includegraphics[width=0.85\textwidth]{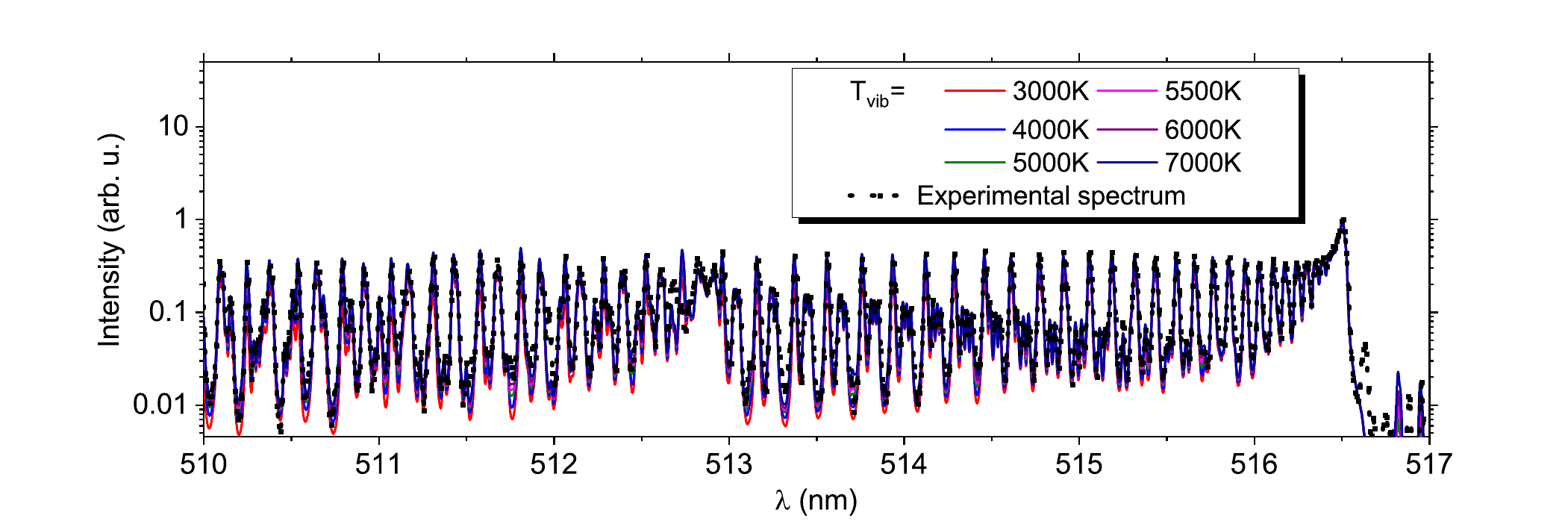}
\includegraphics[width=0.55\textwidth]{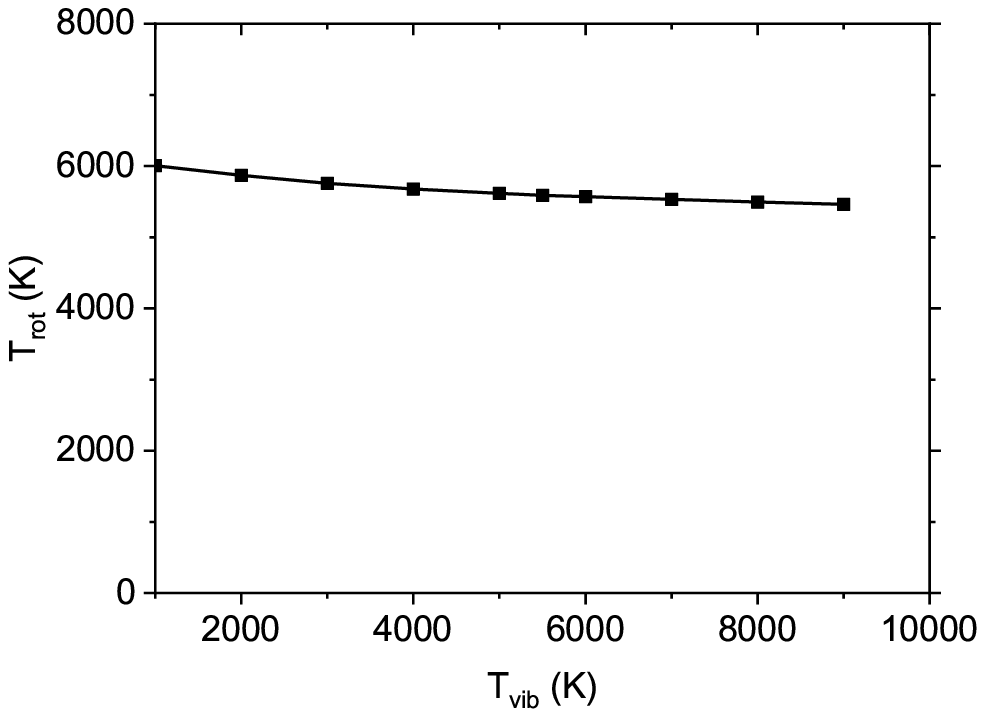}
\caption{Fitting of an experimental C$_2$ $\Delta \nu=0$ Swan band transition group with fixed vibrational temperatures in the spectral range 480-516~nm. On top are superposed spectra normalized to the band head onto the experimental spectrum. On the bottom, are the fitted rotational temperatures while fixing the vibrational temperature. The experimental conditions are the same as in figure \ref{fig:linebyline}.}
\label{fig:C2TvibFixed} 
\end{figure}

Finally, in figure \ref{fig:C2TvibFixed}, the sensitivity of the $\Delta\nu=0$ transition group for determining T$_{rot}$ on assumed/fitted values of T$_{vib}$ is assessed. An experimental spectrum for the $\Delta\nu=0$ transition group is fitted while taking different values for T$_{vib}$. One can see that the fitted rotational temperature is almost independent of the vibrational temperature. This is due to the low transition probabilities for the $\Delta \nu=0$ transitions for the v=1,2 levels while higher levels are even not visible at all. We note that the almost independence of the fitted rotational temperature to $T_{vib}$ is true for the relatively high resolution of the present spectrum. Interdependence of T$_{rot}$ and T$_{vib}$ becomes larger for lower resolution spectra also for the $\Delta \nu=0$ transition group.

The $\Delta\nu=-1,\pm 2$ transition groups usually suffer from low intensities because of their low transition probabilities. The resulting lower signal to noise ratios make these groups not prime candidates for temperatures determination. From figure \ref{fig:sensitivity_contour_theo} and \ref{fig:sensitivity_contour} and the discussion before, it can be concluded that the C$_2$ Swan $\Delta\nu=0$ transition group is the best candidate for the determination of T$_{rot}$ while  $\Delta\nu=\pm1$ should be used for the determination of T$_{vib}$. We should however stress that the present \lq\lq error\rq\rq calculations made here cannot be generalized to any C$_2$ Swan band fitted spectrum but the trends will remain similar. 

\section{Is C$_2$ Swan band emission a good thermometer?}
\label{C2relax}

Before presenting a detailed analysis of the gas temperature of the plasma torch, a comparative study of the temperatures of the C$_2$ Swan band and CN violet system is presented in section \ref{Subsec:CN}. The good agreement that is reported between these two species gives a hint that the temperatures measured are with their local environment and so correspond to the gas temperature T$_{gas}$ with a value of about 6000~K. For a similar plasma source and operating conditions, Groen et al \cite{Groen2019} obtained similar values of 6000~K via 777 nm O line broadening. On the same device, den Harder et al \cite{denHarder2017} estimated the gas temperature by Rayleigh scattering of 5000~K at 200~mbar. Interestingly, for a N$_2$ plasma, Gatti et al \cite{Gatti2018N2}	obtained by Raman scattering on N$_2$ gas temperatures of 7000~K near atmospheric pressure. At atmospheric pressure, we also measured the first negative and second positive system of N$_2$ while operating the present plasma torch in pure N$_2$ and obtained rotational and vibrational temperatures in the order of 6500~K \cite{carbone2018optical}. The consistency in the obtained values  (for high pressure, high power microwave plasmas) using different measurements methods are a first indication that the C$_2$ Swan band may also be used for gas temperature determination.

As discussed in section \ref{C2kin}, there are several pathways involving neutral species for the formation of the C$_2 (d^3\Pi_g$) state. Depending of its formation mechanism, one can expect the nascent rovibrational distribution of an excited state formed by chemical reactions to follow a thermal distribution or to be strongly non-thermal (i.e. formation of the excited into specific rotational and or vibrational state(s). In the latter case, for being able to predict its emission spectrum, it is necessary to take into account collisional relaxation processes \cite{brockhinke2012}. The rotational and vibrational temperatures do not then correspond to the gas temperature. In sections \ref{sec:Vibrelax} and \ref{sec:Rotrelax} are discussed the collisional relaxation kinetics of the C$_2 (d^3\Pi_g$) state. These processes that lead to the thermalization of the C$_2$ (d$^3\Pi_g$) state rotational and vibrational distribution functions define the cases when the C$_2$ Swan band emission spectrum can still be used to measure the gas temperature independently of the formation mechanism of the C$_2$ (d$^3\Pi_g$) state.




\subsection{Comparison with the CN violet system}
\label{Subsec:CN}

By adding a small amount of nitrogen, one can observe a strong decrease of the Swan band emission intensity while the CN violet system increase drastically. Vacher, Lino da Silva et al. \cite{Vacher2008, LinoSilva2008} studied also the effect of adding N$_2$ to a CO$_2$ inductively coupled atmospheric pressure plasma and observed similar trends. Such behavior was also seen by other authors in microwave plasmas and low and atmospheric pressure \cite{TIMMERMANS1999, TIMMERMANS1998}. This tendency can be explained by competing mechanisms for the formation of C$_2$ and CN that goes in favor of the formation of CN with increasing N$_2$ addition and for which it was found that C atoms are the main precursor \cite{DONG201462}. CN (B-X) violet system is often used as a molecular pyrometer for determination of gas temperature in N$_2$ plasma sources with hydrocarbons \cite{Bruggeman2014RotT, Nassar2012, Moon2009, ricard1995torche}. In order to generate CN violet system emission, 0.3 L/min of N$_2$ was added to the plasma corresponding to 3\% admixture. This is a small quantity that, in the case of a pure CO$_2$ plasma, does not perturb the electron properties and neutral particles energy balance of the plasma.

\begin{figure}[ht]
\centering
\includegraphics[width=1\textwidth]{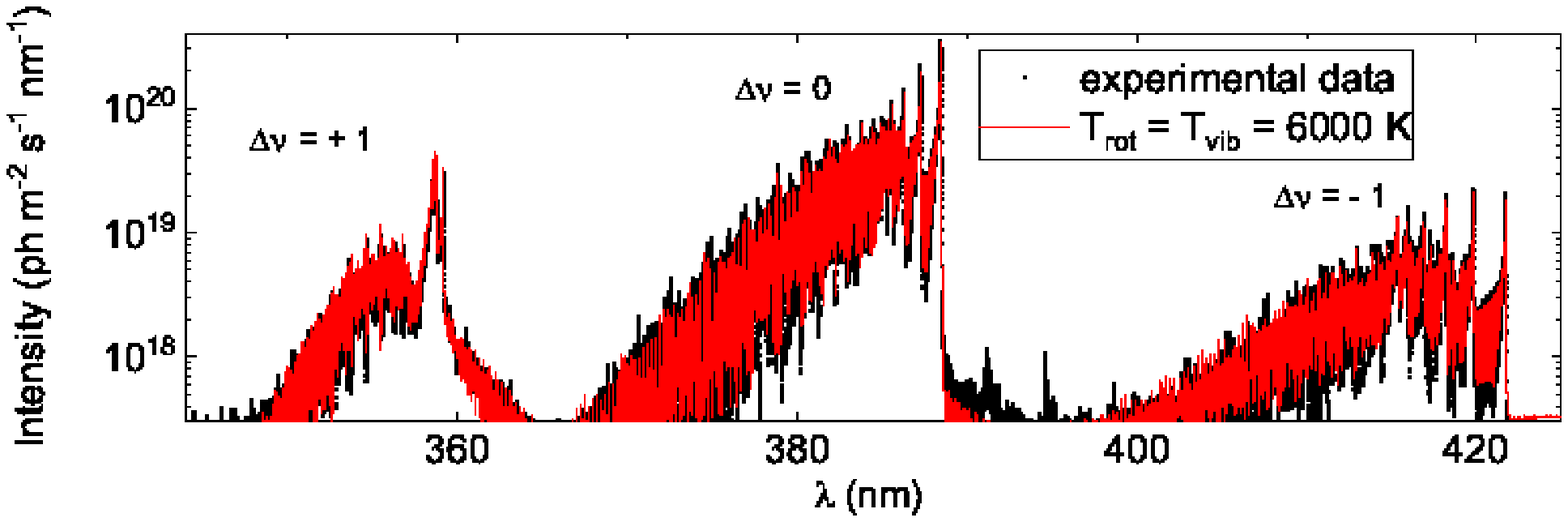}
\caption{CN violet system emission between 345~nm and 425~nm corresponding to a typical temperature of  T$_{rot}\approx$ T$_{vib}$ 6000 K. The spectrum was acquired in the center of the resonator for a power of 900 W and a flow of 10~L/min of CO$_2$ and 0.3~L/min of N$_2$.}
\label{fig:CN_fit} 
\end{figure}

\begin{figure}[ht]
\centering
\includegraphics[width=0.7\textwidth]{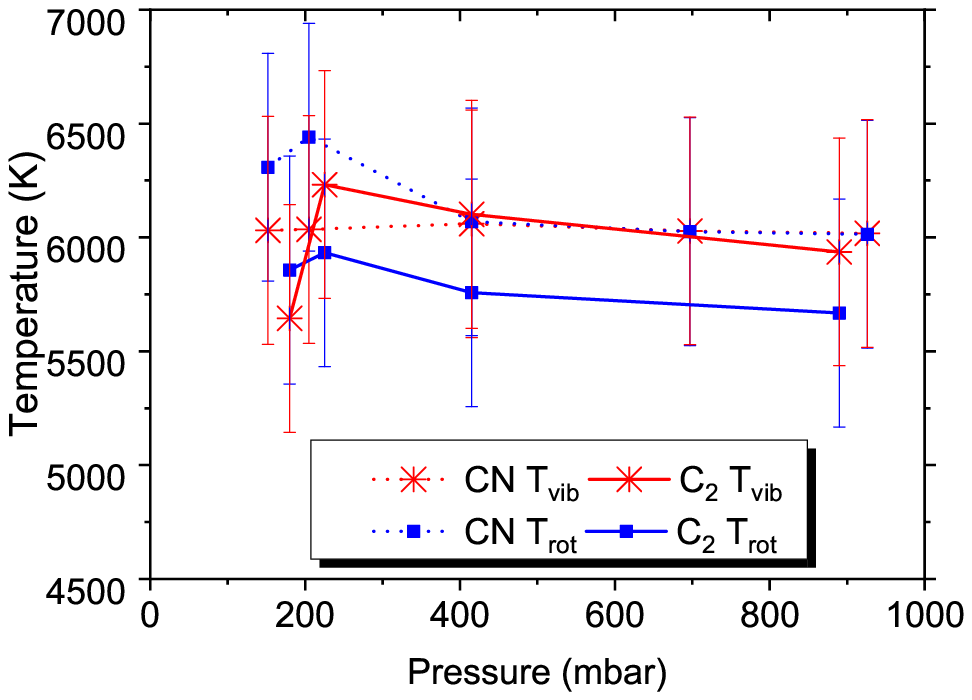}
\caption{Comparison of the rotational and vibrational temperatures of the C$_2$ Swan band obtained by fitting the $\Delta \nu = -1, 0, +1$  transition groups and the CN violet band (fitting of $\Delta \nu =-1, 0, +1$  transition groups) as a function of pressure. The measurements are performed in the center of the resonator using 10 L/min CO$_2$ gas flow and 900 W input power. N$_2$ added gas flow is 0.3~L/min.}
\label{fig:C2_vs_CN}
\end{figure}

In an argon atmospheric pressure microwave plasma with nitrogen addition and CO$_2$ advection from ambient air, Ridenti and Amorim discussed the weak presence of the CN violet system that is strongly overlapping with nitrogen first negative and second positive systems \cite{Ridenti2018CN}. The band head of the first negative system is well separated from the one of the CN violet system while N$_2$ second positive system has distinctive features that do not overlap either with the CN violet system. These two systems can therefore, using high resolution spectroscopy, easily be identified. Their presence was not detected in the present CO$_2$ microwave plasma with 3\% N$_2$ addition. They are however observed in the present reactor while operated at atmospheric pressure with pure N$_2$ flow \cite{carbone2018optical}. In figure \ref{fig:CN_fit}, an example of CN spectrum and its resulting fitting are shown. A value of T$_{rot}$=T$_{vib}=6000 \pm 500$~K is obtained which are the same as for the Swan band system. Also the residue of the fit shows that no deviation from rotational or vibrational Boltzmann distribution can be observed experimentally. A comparison for several pressures between the measured rotational and vibrational temperatures obtained by fitting the two molecular bands systems is presented in figure \ref{fig:C2_vs_CN}. One can see that, within error bars, there are no significant differences for the two molecular systems and that the mean values agree between each other very well. Note that in the case of CN, also the rotational and vibrational temperatures were fitted independently (cf. method discussed in section \ref{StatebState}). This confirms that the gas inside the plasma region where the light comes from is very hot with a temperature of about 6000~K both for rotational and vibrational internal degrees of freedom. The rovibrational distributions for those two systems can be essentially described with a single temperature.

\subsection{Collisional relaxation of vibrational states}
\label{sec:Vibrelax}

C$_2$ Swan bands emission during pulsed laser irradiation of soot or graphite is often used for determination of gas temperature in (sub-)microsecond and even nanosecond time scales (see for instance \cite{Saito2003, Goulay2010}). The spectral resolution of such spectra is however usually low (because of the relative low intensity of such discharges) and a detailed evalution of the thermalization of rotational and vibrational states is difficult. UV pulsed photolysis experiments by Faust et al. \cite{faust1981time} showed that the formation of the high pressure band is occuring only on long time scales (t$>>1\mu$s) while for time up to 1$\mu$s the Swan band emission presented a normal vibrational distribution structure. They concluded that two different processes contribute to the formation of the C$_2$ Swan bands and that vibrational relaxation of the v=6 level is inefficient for populating the v=0 level of the d$^3\Pi_g$ state.



Bondybey \cite{Bondybey1976} analyzed the C$_2$ Swan band fluorescence following two photon excitation in rare gas solids. He observed an erratic progression of lifetimes for different vibrational levels and concluded that the fast vibrational relaxation of the $d^3\Pi_g$ v=3,4 states may occur through the $b^3\Sigma_g^-$ state, following a suggestion by Frosch \cite{Frosch1971}. Xuechu and Nanquan \cite{xuechu1980vibrational} studied the vibrational relaxation of the $C_2 (d^3\Pi_g)$ state by observing the chemiluminescence spectrum of the reaction Na + CCl$_4$ in a thermal bath of argon where the $C_2 (d^3\Pi_g, v=6)$ state is formed preferentially. From their pressure dependence study, they concluded that the thermal relaxation of the  $C_2 (d^3\Pi_g)$ state by argon takes only a few collisions with a rate of $2.2\cdot 10^6$Torr$^{-1}$s$^{-1}$ while they deduce even higher quenching rate by the sodium atoms. 

Several experiments have shown that vibrational thermalization of the C$_2$ (d$^3\Pi_g$) state is relatively slow and potentially as long if not longer than the lifetime of the state itself excepted for the rate reported above by Xuechu and Nanquan \cite{xuechu1980vibrational}. It may therefore be surprising to see that the vibrational distribution function follows closely a Maxwell-Boltzmann distribution in the experiments reported in this paper. However, the present measurements are done at high temperatures (i.e. $T>$ 1000~K) and one may expect that the nascent population distribution follows already closely a thermal distribution, partly because of the large translational motion of the reactants. A temperature of 6000~K corresponds to a kinetic energy of 0.5~eV which is more than the energy separation between individual vibrational levels. 

Rich and Bergman \cite{rich1979c2} performed vibrational excitation of CO at room temperature using a CO laser and observed formation of the $C^1\Pi_g$ and $d^3\Pi_g$ states of C$_2$. They claimed that the observation of the $C^1\Pi_g$ state (at the origin of the $C^1\Pi_g$-$A^1\Pi_u$ Deslandres-D'Azambuja system) was the indication that CO (v) states can electronically excite the C$_2$ molecule. Indeed, the $C^1\Pi_g$ state can be formed only via the association of two C ($^1$D) states and in their experiment only C ($^3$P) ground state atoms were produced. However, an analysis of the potential energy curves of C$_2$ shows that the $C^1\Pi_g$ has an avoided crossing with another $^1\Pi_g$ state and consequently dissociates into two ground state C($^3$P) atoms \cite{ballik1963extension}. Grigorian and Cenian \cite{grigorian2009vibrational} recently performed a similar experiment and reported also the formation of the $e^3\Pi_g$ state ($e^3\Pi_g$-$a^3\Pi_u$, Fox-Herzberg system) which have for dissociation limit the C ($^3$P) and C($^1$D) states \cite{martin1992c2}. Moreover, Wallaart et al. \cite{Wallaart1995, WALLAART1995b} reported during a CO vibrational excitation experiment the formation of the D$^1\Sigma_u^+$ state (D$^1\Sigma_u^+$-X$^1\Sigma_g^+$, Mulliken system) in addition to the aforementioned ones.  The D$^1\Sigma_u^+$ state has for dissociation limit  the C ($^1$S)+C($^1$D) states \cite{martin1992c2}. Although the initial experiments of Rich and Bergman did not allow to support their hypothesis of direct electronic excitation of C$_2$ electronic states by CO(v), subsequent measurements confirm their proposal. These observations go in favor to the hypothesis that vibrationally excited CO molecules can directly populate C$_2$ electronically excited states and that multiple vibrational excitation quanta may occur simultaneously. We note that Wallaart et al. reported also the observation of radiative transitions from electronically excited C atoms which could also be excited by CO(v). The presence of C$_2$ (D$^1\Sigma_u^+$) and C$_2$ ($e^3\Pi_g$) states in the aforementioned studies could therefore also be explained by the recombination of C atoms which are electronically excited by CO(v) after being formed by the Boudouard mechanism (cf. reaction \ref{boudouardClassic}). The relatively low density of C* atoms compared to CO(v) states however makes it difficult to assess how important that channel is in reality. Hack and Langel \cite{HACK1981387} analysed the recombination of C($^1$D) atoms and observed the production of C$_2$(e$^3\Pi_g$, C$^1\Pi_g$) electronic states but no recombination rate coefficient was reported in the literature yet. The usual observation of C$_2$ electronic states formation in C/CO mixtures without detection of any electronically excited states of C atoms (at the exception of the work of Wallart et al) tends to support that efficient energy exchange mechanisms between CO(v) and C$_2$ exist. Such processes will favor rapid thermalization of the vibrational states of the C$_2$ molecule. Similarly, one can expect that the CO$_2$ molecule also interact with C$_2$ molecules in similar fashion due to its rich vibrational structure. Based on the present discussion, it can be concluded that either the formation mechanism of the C$_2$ (d$^3\Pi_g$) state has already a nascent (nearly)-thermalized vibrational distribution (which seems unlikely based on its formation mechanisms discussed in section \ref{C2kin}) or that, not only collisional quenching processes, but also reactive energy exchanges leads to its (faster) thermalization. This would explain the invariant observation of thermalized vibrational distributions of the C$_2$ (d$^3\Pi_g$) state in CO$_2$ microwave plasmas.

\subsection{Collisional relaxation of rotational states and gas temperature}
\label{sec:Rotrelax}

Direct measurements of thermalization rates are usually difficult to obtain and indirect estimations are usually done. Using the assumption that the C$_2$ molecule would be formed rotationally hot and that the pressure dependence of the C$_2$ Swan band temperature can be used as estimate of the rotational relaxation/equilibration with the thermal bath of molecules,  Bleekrode determined a rotational relaxation time of 3$\mu$s in an oxyacetylene flame at low pressure (2-120 Torr) \cite{Bleekrode1966}. Using the same hypotheses, Kini and Savadatti \cite{kini1977investigation} estimated a rotational relaxation time for the Swan band of 30~$\mu$s in a CO discharge in a pressure range of 2-24 Torr. The underlying assumptions were particularly strong as no elementary process for the formation of rotationally hot C$_2$ were identified. However, Brockhinke et al \cite{brockhinke2006energy} recently performed picosecond laser induced fluorescence measurements on the $\Delta\nu=0$ transition group of the C$_2$ (d$^3\Pi_g$-a$^3\Pi_u$) Swan. They managed to evaluate rotational energy transfer (RET) and vibrational energy transfer (VET) by monitoring the collisional transfer of single optically pumped rovibrational state of the d$^3\Pi_g$ state to adjacent states by fluorescence. The measurements were performed in an oxy-acetylene flame with a gas temperature of 2200~K. Such measurements allow to monitor individual state to state RET and VET collisional transfers and are therefore very precise. They obtained state-to-state collisional re-distribution rate for RET in the order of $k_{RET}\approx 4\cdot 10^{9}$s$^{-1}$. Due to the high electronic quenching rate of the C$_2$ d$^3\Pi_g$ states they could not quantify the VET rate but assessed that it was slower than $k_{RET}$ in agreement with aforementioned studies. 

The temperatures obtained by fitting the C$_2$ Swan band and the CN (B$^2\Sigma^+$ - X$^2\Sigma^+$) violet system have been compared and match very well as seen in figure \ref{fig:C2_vs_CN}. Contrary to the C$_2$ (d$^3\Pi_g$) state, the CN (B$^2\Sigma^+$) state does not have, to the best of our knowledge, known mechanisms leading to non-equilibrium rovibrational emission spectra \cite{Grigorian2011}. Rotational relaxation times of the CN (B$^2\Sigma^+$) state are also considerably shorter than its radiative lifetime \cite{Guo2000CN, Brunet2002}. For CN, the rotational and vibrational states distributions have time to thermalize before being observed by optical emission spectroscopy in continuous plasmas. The cross check between the rotational temperatures of the Swan band and the CN violet system gives additional confidence that their rotational temperature may be associated to the gas temperature. The reactive formation of molecules can only lead to an overpopulation of the high rotational levels of a molecule (because of energy exchange during the formation process) or to a thermal distribution. In the case of non-Boltzmann distribution of rotational levels, the temperature of the lower rotational levels is then usually taken as a measure of the gas (i.e. translational) temperature of the plasma \cite{Bruggeman2014RotT}. In the present study (see figure \ref{fig:statebstateAll}), two-temperature rotational distribution function are not found.	Observing thermal distributions of rotational states of the C$_2$  (d$^3\Pi_g$) state and knowing that its RET relaxation time is much shorter than its radiative lifetime gives confidence that T$_{rot}$ is a local measure of the translational temperature of the plasma. This is additionally supported by the equivalence between the rotational and vibrational temperatures both for the C$_2$ Swan bands and for the CN violet system. For high pressure CO$_2$ plasmas (i.e. typically 100~mbar and above), it can therefore be proposed to use the C$_2$ Swan bands rotational temperature as an evaluation tool for the local gas temperature of the plasma. We note however, that for a CO$_2$ plasma with a much lower gas temperature, similar analysis should be performed as in section \ref{StatebState}.

\section{Parametric analysis of the CO$_2$ microwave plasma torch}
\label{sec:results}

Having assessed that C$_2$ Swan band can be used for measuring the gas temperature in a CO$_2$ plasma, we report in this section a parametric study of C$_2$ T$_{rot}$ and T$_{vib}$ as a function of plasma parameters. Although both T$_{rot}$ and T$_{vib}$ can be associated to the gas temperature, they are reported here independently. The measurements are taken in the pressure regime above 120 mbar where the C$_2$ Swan band emission is the dominant feature in the plasma emission spectrum. Concomitantly, the plasma is contracted with a diameter of about 3-8 mm which depends mainly of the applied power. The emission is spatially discriminated using a system of two pin-holes and a lens coupled to a fiber and collected over a area of about 1~mm$^2$. The latter corresponds to the spatial resolution attainable in the present setup. Each data point corresponds to the light collected along the light of sight (radial direction) and over a plasma slab of 1~mm in the axial direction. Interestingly, in the contracted regime of the plasma torch (i.e. above a pressure of about 120 mbar), axial and radial measurements of the C$_2$ rotational temperature show no variation for all studied plasma conditions. After substraction of the continuous baseline, only a spatial variation of the overall intensity of the C$_2$ Swan band is observed within signal to noise sensitivity. In the following, all data are consequently reported for measurements in the center of the resonator; i.e 20~mm above the bottom of the waveguide while looking across the axis of the quartz tube. 


\begin{figure}[ht]
\centering
\includegraphics[width=0.7\textwidth]{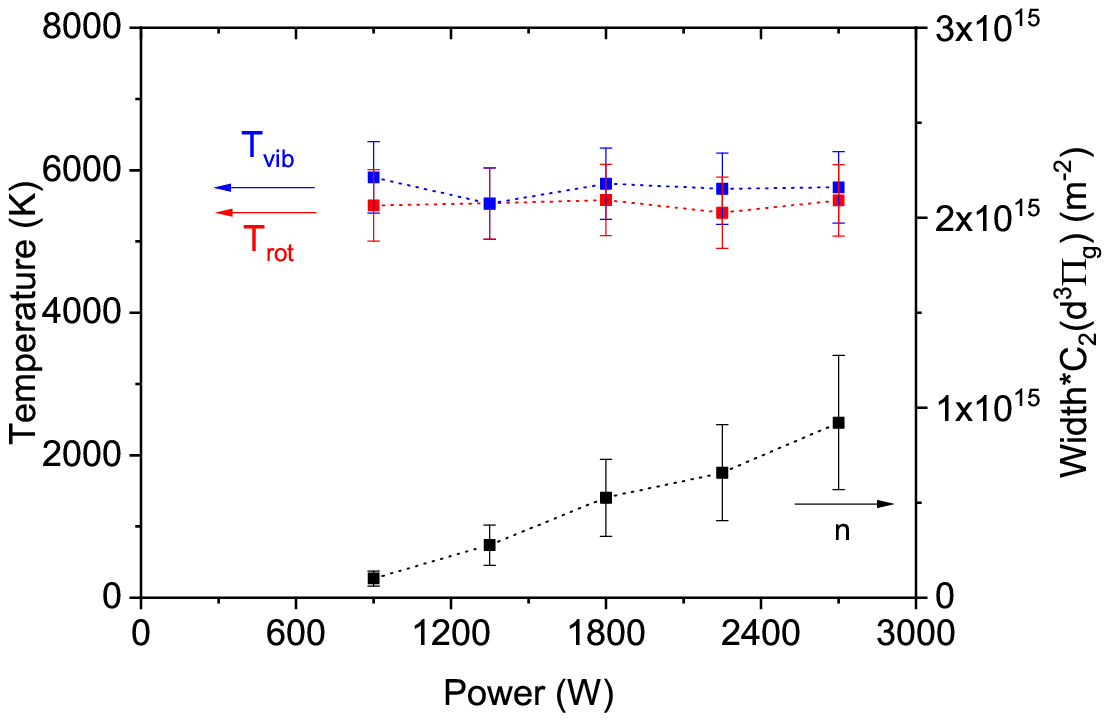}
\caption{Rotational and vibrational temperature of the C$_2$ Swan band obtained by fitting the $\Delta \nu =0$ and $\Delta \nu =0, \pm1$ transition groups respectively as a function of power input. The measurements are performed in the center of the resonator at 925 mbar with 10 L/min CO$_2$ gas flow. $n$ is the laterally integrated density of the C$_2$($d^3\Pi_g$) state.}
\label{fig:C2_atm_power}
\end{figure}

In figure \ref{fig:C2_atm_power}, the effect of the input microwave power on the rotational and vibration temperatures of the C$_2$ Swan bands are presented. The measurements are done near atmospheric pressure plasma (925 mbar) for a gas flow of 10 L/min. One can see that both the rotational and vibrational temperatures of C$_2$ remain constant and equal within error bars. The increase in power has only the consequence of increasing the intensity of the radiation from the C$_2$($d^3\Pi_g$) state. This increase of intensity is correlated with an increase in plasma radius. Also, by imaging it is observed that the plasma becomes elongated in the effluent region with increasing power.

\begin{figure}[ht]
\centering
\includegraphics[width=0.7\textwidth]{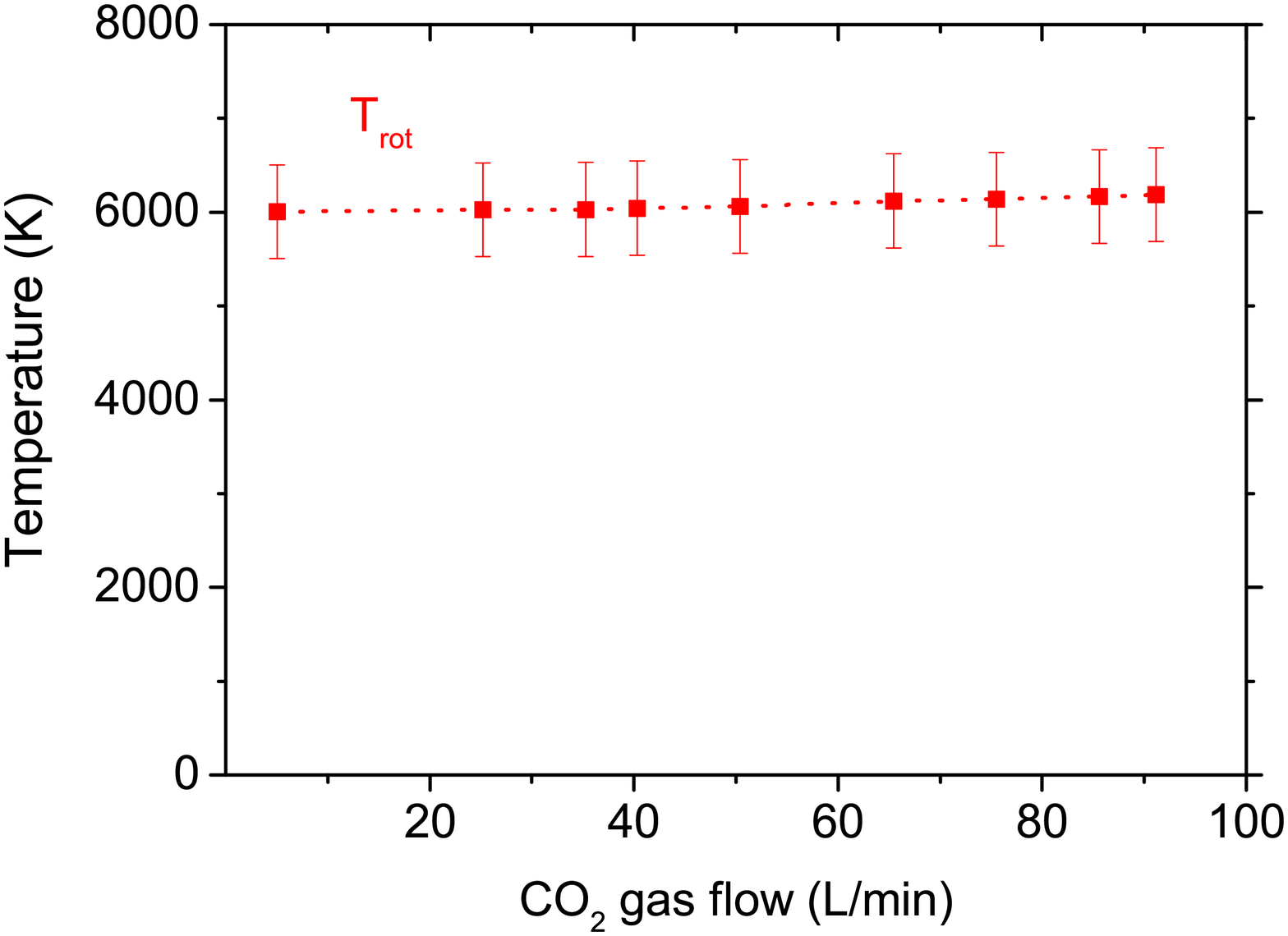}
\caption{Rotational/gas temperature of the C$_2$ Swan band obtained by fitting the $\Delta \nu =0$ transition group as a function of the gas flow. The measurements are performed in the center of the resonator for a pressure of 850 mbar and an input power of 900 W.}
\label{fig:C2_gasflow}
\end{figure}

In order to cool down a high pressure plasma, one possible way is by increasing the gas flow. In figure \ref{fig:C2_gasflow}, the variation of the C$_2$ rotational temperature versus CO$_2$ gas flow at a pressure of about 850~mbar is shown for an input power of 900~W. One can see that the temperature remains constant around 6000$\pm$ 500~K. Such trend was also observed by Sun et al. \cite{SunHojoong2017} and Spencer and Gallimore \cite{Spencer2013MWcat} but for lower flow rates variation. Similar lack of influence of the gas flow was also observed in the case of an atmospheric pressure microwave plasma torch in nitrogen \cite{ChuanJieChen2015}. Increasing the gas flow by more than 10-fold appears in the present setup (and also for other high pressure microwave sources) to be an inefficient way of tuning the gas temperature despite reducing the specific energy input per molecule by the same factor. Such result is rather surprising but may be due to an inefficient mixing between the hot and cold gas region.

\begin{figure}[ht]
\centering
\includegraphics[width=0.7\textwidth]{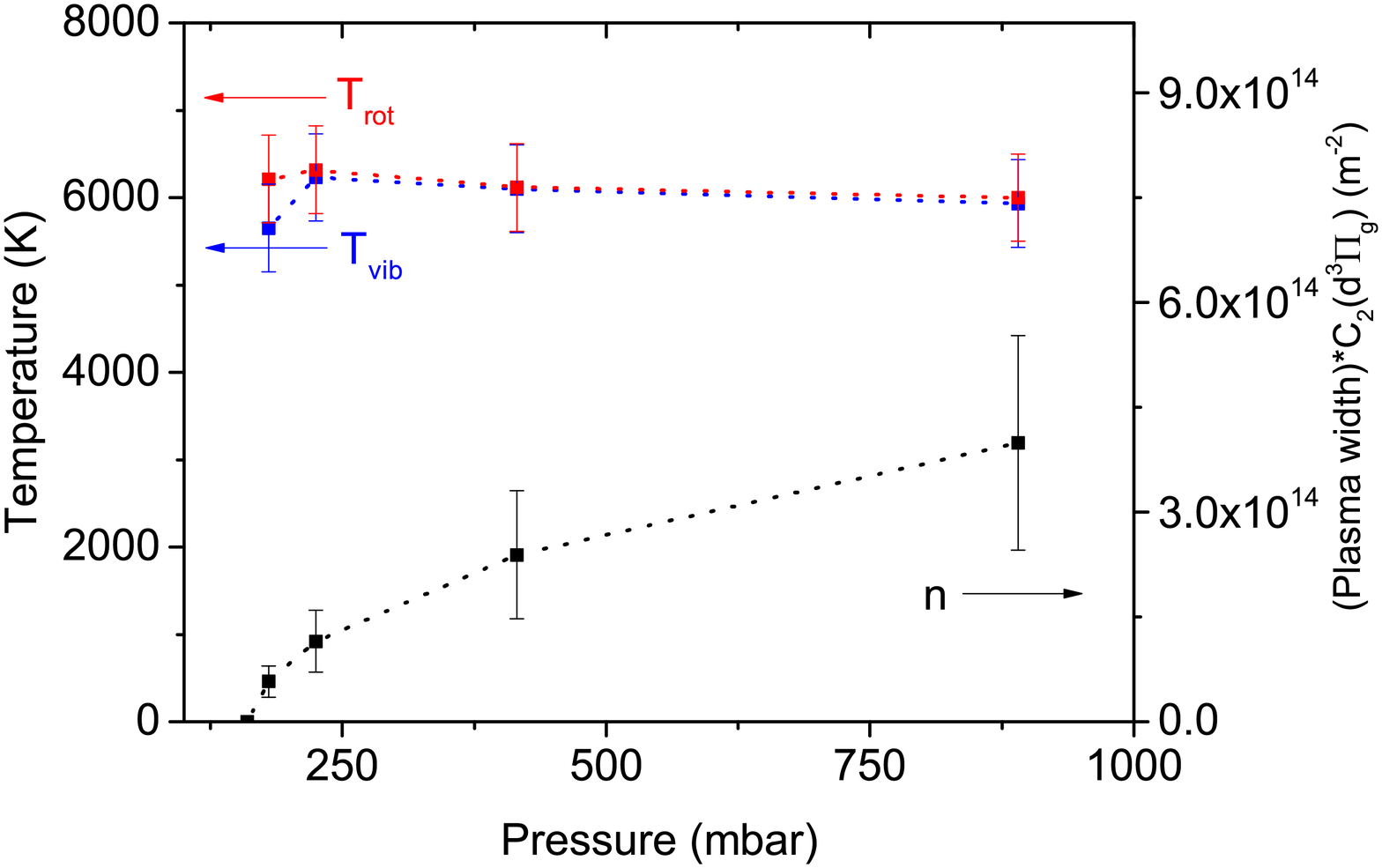}
\caption{Rotational and vibrational temperature of the C$_2$ Swan band obtained by fitting the $\Delta \nu =0$ and $\Delta \nu =0, \pm1$ transition groups respectively as a function of pressure. The measurements are performed in the center of the resonator using a 10 L/min CO$_2$ gas flow and 900 W input power.}
\label{fig:C2_pressure}
\end{figure}

A typical method for enhancing the non-equilibrium of a plasma discharge (e.g. reducing the gas temperature while the electron temperature remains in the order of 1~eV or more) is to reduce the gas pressure. In figure \ref{fig:C2_pressure}, the pressure is varied from near atmospheric pressure down to 180 mbar where the C$_2$ Swan band is still intense enough for getting a good signal to noise ratio. One can see that $T_{vib}$ and $T_{rot}$ are equal to each other within absolute uncertainties. One should note that in pure CO$_2$ microwave plasmas, lower temperatures have been measured but only at (much) lower pressures than in the current study and where C$_2$ Swan band is not observed \cite{TIMMERMANS1999, Silva2014PSST}. On the other hand, den Harder et al. \cite{denHarder2017} and Groen et al \cite{Groen2019} measured gas temperature in the order of 5000 to 6000~K in the same pressure range as for the present study. 


\section{Conclusions and outlook}

In this paper, the potential sources of C$_2$ molecules in CO$_2$ plasmas and the known pathways leading to non-equilibrium vibrational distributions of the $(d^3\Pi_g)$ state are reviewed. A detailed analysis of high resolution spectra is performed in order to detect potential deviations of the rovibrational distribution functions from thermal equilibrium in a high pressure microwave plasma discharge. State of the art molecular constants were used in order to fit the spectra. It is found that Boltzmann distributions can be used both for rotational and vibrational states for all conditions in the present study. Consequently, potential source terms of the C$_2$ $(d^3\Pi_g)$ state in a CO$_2$ plasma do not lead to any observed deviation from a Boltzmann vibrational distribution function. The sensitivity of different parts of the spectra for determining T$_{rot}$ and T$_{vib}$ is evaluated. For a fast and accurate determination of the rotational temperature of the C$_2$ $(d^3\Pi_g)$ state, it is shown that the $\Delta\nu =0$ transition group is the most accurate part of the spectrum  to be taken into account. 

Due to fast thermalization of the rotational states of the C$_2 (d^3\Pi_g)$ state with the background gas, the measured T$_{rot}$ can be used for assessing the gas temperature of the plasma. Furthermore, the vibrational states of the C$_2 (d^3\Pi_g)$ level are found to be thermalized indicating the possible presence of fast thermalization processes. 

A parametric study is performed on a pure CO$_2$ microwave plasma torch and the both the vibrational and rotational temperatures are found to be insensitive to external plasma parameters such as gas flow and input power at pressures of 180~mbar and above and a value of 6000 $\pm$ 500 K is found. The latter can be associated to the gas temperature of the plasma. These results are in agreement with the hypothesis that chemical equilibrium of the species can be described by the gas temperature at which thermal dissociation play an important role. The only observable differences in the extended parameter range of the present study are in the radially integrated density of the C$_2$ (d$^3\Pi_g$) state. We note that similar values as the one presented in this study were reported before in a certain number of studies but no assessment of non-equilibrium distributions was done. Also, the rotational temperature of C$_2$ was implicitely assumed previously to be the one of the gas temperature and it is now shown that such approximation in high pressure CO$_2$ microwave discharges appears to hold.

The  C$_2$ (d$^3\Pi_g$)  state emits only in a narrow region in the center of the plasma and only the gas temperature in the core of the plasma can be measured. The region inside the plasma reactor that can be probed is only where C atoms are present and react to form C$_2$ (or CN when some N$_2$ is present) molecules. It was observed that C atoms are not observed by optical emission spectroscopy at pressures lower than the ones where the C$_2 (d^3\Pi_g)$ state is measured. To form carbon atoms, large power densities (i.e. high T$_{vib}$ and/or T$_{gas}$) are required as CO is the only species that can act as a primary source of carbon atoms. Interestingly, at the observed temperature of 6000~K, CO can already partially thermally dissociate \cite{Fairbairn1969}. This implies that observation of C and C$_2$ species is linked to regions with high gas temperatures where CO$_2$ molecules are readily dissociated. 

The measurements reported in this paper show the presence of a hot core of gas but no insight in the temperature gradient at the edge of the plasma could be obtained as not radiative emitting molecular species can be observed at the edges of this hot core. High gas temperatures T$\geq$ 3500K allow already to dissociate fully the CO$_2$ and partially the CO molecules through thermal processes \cite{Bekerom2018, Fridman2008}. \textit{In situ} diagnostics techniques like Raman scattering and CARS \cite{Lempert_2014, vandenBekerom18OSA} would bring additional understanding in the gas dynamics at the boundary between the hot plasma core and the colder edge.

\section*{Acknowledgments}

The authors would like to acknowledge prof. Christophe Laux (Ecole Centrale, Paris) for giving access to his program Specair. 
Fruitful discussion and collaboration with our colleagues from Uni. Stuttgart (Irina Kistner, dr. Andreas Schulz, dr. Sandra Gaiser, dr. Matthias Walker), DIFFER (dr. Tiny Verreycken, dr. Floran Peeters, dr. Waldo Bongers) and KIT (Lucas Silberer, dr. Sergey Soldatov) are also gratefully acknowledged.

\vspace{1cm}

\bibliographystyle{unsrt}

\bibliography{C2spectro_bib}

\newpage
\begin{appendices}

\section{Molecular constants for the calculations of the C$_2$ Swan band system}
\label{C2appendix}

\begin{table}[ht]
\resizebox{\textwidth}{!}{%
\begin{tabular}{|l|l|l|l|l|l|l|l|l|l|}
\hline
$\nu$  & Origin & A& A$_D$ & B$_\nu$ & $10^6 \cdot $ D$_\nu$ & $\lambda$ & o & p & q  \\
\hline
0  & 19 378.46749(51) & -14.00139(63) & 0.0005068(83)  & 1.7455663(43) & 6.8205(16) & 0.03301(47)  & 0.61076(52) & 0.003973(43) & -0.0007752(43) \\
1  & 21 132.14977(25) & -13.87513(49) & 0.0005740(83)  & 1.7254062(53) & 7.0194(77) & 0.02972(38)  & 0.61713(36) & 0.004133(44) & -0.0008171(43) \\
2  & 22 848.3877(21)  & -13.8205(23)  & 0.000600(43)   & 1.704516(21)  & 7.308(22)  & 0.0253(41)   & 0.6208(32)  & 0.00624(38)  & -0.000835(14)  \\
3  & 24 524.2201(19)  & -13.5361(28)  & 0.000775(17)   & 1.681437(16)  & 7.438(24)  & 0.0470(26)   & 0.5827(26)  & 0.00579(17)  & -0.0008568(85) \\
4  & 26 155.0448(29)  & -13.3892(50)  & 0.001451(14)   & 1.656859(26)  & 7.684(43)  & 0.0219(38)   & 0.6313(32)  & 0.00954(29)  & -0.000923(21)  \\
5  & 27 735.6720(43)  & -13.0324(66)  & 0.000723(37)   & 1.630205(23)  & 8.573(32)  & 0.0601(28)   & 0.6161(23)  & 0.00685(32)  & -0.000912(15)  \\
6  & 29 259.3548(36)  & -12.820(10)   & 0.001203(56)   & 1.599876(31)  & 8.998(44)  & 0.0529(71)   & 0.5773(71)  & 0.00874(47)  & -0.000986(22)  \\
7  & 30 717.9011(46)  & -12.3458(71)  & 0.000814(41)   & 1.566047(32)  & 10.044(66) & 0.0960(34)   & 0.5532(31)  & 0.00936(35)  & -0.001175(17)  \\
8  & 32 102.655(22)   & -12.107(22)   & 0.00076(fixed) & 1.52675(31)   & 9.60(97)   & 0.095(fixed) & 0.546(22)   & 0.0055(21)   & -0.00088(21)   \\
9  & 33 406.230(22)   & -11.698(39)   & 0.00076(fixed) & 1.485755(96)  & 11.85(10)  & 0.172(26)    & 0.498(28)   & 0.0097(14)   & -0.002062(44)  \\
10 & 34 626.7860(94)  & -11.297(15)   & 0.00076(fixed) & 1.441138(72)  & 12.837(73) & 0.115(16)    & 0.399(12)   & 0.00745(90)  & -0.000977(30) \\
\hline
\end{tabular}}
\caption{Molecular constants (in cm$^{-1}$) for the $d^3\Pi_g$ electronic state used in the calculations performed by Brooke et al. \cite{BROOKE201311}. The uncertainties are given in parenthesis (i.e. one standard deviation to the last signiﬁcant digits of the constants). For the vibrational level $\nu$=1 the additional constant H$=2.14(30) \cdot 10^{-11}$ is used.}
\label{tab:C2constant}

\end{table}
\begin{table}[ht]

\resizebox{\textwidth}{!}{%
\begin{tabular}{|l|l|l|l|l|l|l|l|l|l|}
\hline
$\nu$  & Origin & A& A$_D$ & B$_\nu$ & $10^6 \cdot $ D$_\nu$ & $\lambda$ & o & p & q  \\
\hline
0 & 0               & -15.26986(43) & 0.0002634(71)  & 1.6240452(44) & 6.4506(19) & -0.15450(36) & 0.67525(35) & 0.002537(42) & -0.0005281(44) \\
1 & 1618.02244(53)  & -15.25197(61) & 0.0002266(73)  & 1.6074266(44) & 6.4439(21) & -0.15373(51) & 0.67017(51) & 0.002705(44) & -0.0005772(42) \\
2 & 3212.72793(96)  & -15.2328(15)  & 0.0001996(94)  & 1.5907513(61) & 6.4527(44) & -0.1526(12)  & 0.6649(14)  & 0.003132(77) & -0.0006457(48) \\
3 & 4784.0688(31)   & -15.1972(39)  & 0.000186(42)   & 1.574088(24)  & 6.455(24)  & -0.1333(61)  & 0.6815(51)  & 0.00488(42)  & -0.000618(17)  \\
4 & 6332.1364(51)   & -15.2043(65)  & 0.000318(36)   & 1.557117(31)  & 6.338(39)  & -0.1551(72)  & 0.6674(67)  & 0.00632(36)  & -0.000894(16)  \\
5 & 7856.8175(32)   & -15.2096(35)  & 0.00025(fixed) & 1.540139(24)  & 6.312(35)  & -0.1492(36)  & 0.6546(37)  & 0.00734(25)  & -0.001246(12)  \\
6 & 9358.1565(40)   & -15.1646(60)  & 0.000355(33)   & 1.523439(26)  & 6.034(38)  & -0.1551(46)  & 0.6886(38)  & 0.00504(32)  & -0.000676(16)  \\
7 & 10 836.1430(92) & -15.085(11)   & 0.00025(fixed) & 1.50869(25)   & 3.51(92)   & -0.1641(90)  & 0.704(14)   & -0.0244(23)  & 0.00632(49)    \\
8 & 12 290.7997(29) & -15.1702(46)  & 0.00025(fixed) & 1.488684(28)  & 5.329(52)  & -0.1665(36)  & 0.6742(30)  & 0.01449(29)  & -0.002053(23)  \\
9 & 13 722.0897(43) & -15.0980(61)  & 0.000419(31)   & 1.472818(24)  & 6.066(35)  & -0.1584(27)  & 0.6926(23)  & 0.00303(33)  & -0.000081(16) \\
\hline
\end{tabular}}
\caption{Molecular constants (in cm$^{-1}$) for the $a^3\Pi_u$ electronic state used in the calculations performed by Brooke et al. \cite{BROOKE201311}. The uncertainties are given in parenthesis (i.e. one standard deviation to the last signiﬁcant digits of the constants). For the vibrational level $\nu$=0 the additional constants H$ = 6.73(16)\cdot 10^{-11}$, o$_D = -6.86 (114) \cdot 10^{-6}$ and q$_D= -9.61 (41) \cdot 10^{-9}$ are used.}
\label{tab:C2constantII}

\end{table}

\begin{table}[ht]

\resizebox{\textwidth}{!}{%
\begin{tabular}{|l|l|l|l|l|l|l|l|l|l|l|l|l|}
\hline
\multicolumn{2}{|c|}{\multirow{2}{*}{\rotatebox{25}{$\nu''-\nu'$}}}& \multicolumn{11}{c|}{$d^3\Pi_g$}\\
\cline{3-13}

\multicolumn{2}{|c|}{}     & 0          & 1          & 2          & 3          & 4          & 5          & 6          & 7          & 8          & 9	      &10       \\
\hline
\parbox[t]{2mm}{\multirow{10}{*}{\rotatebox[origin=c]{90}{$a^3\Pi_u$}}}
&0 & 7.626 $10^{+6}$ & 2.814 $10^{+6}$ & 2.809 $10^{+5}$ & 4.333 $10^{+3}$ & 2.033 $10^{+2}$ & 3.642 $10^{+1}$ & 2.470 $10^{-2}$ & 2.140 $10^{-1}$ & 9.989 $10^{-4}$ & 3.827 $10^{-3}$ & 3.140 $10^{-9}$ \\
&1 & 2.135 $10^{+6}$ & 3.427 $10^{+6}$ & 4.072 $10^{+6}$ & 6.429 $10^{+5}$ & 8.720 $10^{+3}$ & 1.608 $10^{+3}$ & 1.591 $10^{+2}$ & 4.744 $10^{+0}$ & 1.822 $10^{+0}$ & 4.947 $10^{-2}$ & 3.540 $10^{-2}$ \\
&2 & 3.832 $10^{+5}$ & 2.746 $10^{+6}$ & 1.270 $10^{+6}$ & 4.422 $10^{+6}$ & 9.615 $10^{+5}$ & 7.432 $10^{+3}$ & 6.154 $10^{+3}$ & 3.223 $10^{+2}$ & 6.379 $10^{+1}$ & 4.108 $10^{+0}$ & 1.567 $10^{+0}$ \\
&3 & 5.590 $10^{+4}$ & 8.273 $10^{+5}$ & 2.568 $10^{+6}$ & 3.236 $10^{+5}$ & 4.301 $10^{+6}$ & 1.168 $10^{+6}$ & 1.085 $10^{+3}$ & 1.805 $10^{+4}$ & 1.707 $10^{+2}$ & 3.472 $10^{+2}$ & 1.346 $10^{-2}$ \\
&4 & 7.224 $10^{+3}$ & 1.710 $10^{+5}$ & 1.169 $10^{+6}$ & 2.066 $10^{+6}$ & 2.505 $10^{+4}$ & 4.005 $10^{+6}$ & 1.149 $10^{+6}$ & 4.993 $10^{+3}$ & 3.581 $10^{+4}$ & 1.535 $10^{+2}$ & 9.452 $10^{+2}$ \\
&5 & 8.592 $10^{+2}$ & 2.886 $10^{+4}$ & 3.215 $10^{+5}$ & 1.352 $10^{+6}$ & 1.510 $10^{+6}$ & 1.513 $10^{+4}$ & 3.459 $10^{+6}$ & 1.120 $10^{+6}$ & 5.272 $10^{+4}$ & 4.975 $10^{+4}$ & 4.445 $10^{+3}$ \\
&6 & 9.574 $10^{+1}$ & 4.280 $10^{+3}$ & 6.793 $10^{+4}$ & 4.745 $10^{+5}$ & 1.381 $10^{+6}$ & 1.038 $10^{+6}$ & 8.028 $10^{+4}$ & 3.438 $10^{+6}$ & 8.558 $10^{+5}$ & 1.863 $10^{+5}$ & 4.075 $10^{+4}$ \\
&7 & 1.006 $10^{+1}$ & 5.775 $10^{+2}$ & 1.218 $10^{+4}$ & 1.218 $10^{+5}$ & 6.009 $10^{+5}$ & 1.302 $10^{+6}$ & 6.443 $10^{+5}$ & 1.343 $10^{+5}$ & 3.253 $10^{+6}$ & 4.834 $10^{+5}$ & 4.111 $10^{+5}$ \\
&8 & 9.931 $10^{-1}$ & 7.206 $10^{+1}$ & 1.945 $10^{+3}$ & 2.582 $10^{+4}$ & 1.830 $10^{+5}$ & 6.840 $10^{+5}$ & 1.087 $10^{+6}$ & 4.546 $10^{+5}$ & 1.285 $10^{+5}$ & 3.081 $10^{+6}$ & 1.342 $10^{+5}$ \\
&9 & 9.005 $10^{-2}$ & 8.335 $10^{+0}$ & 2.828 $10^{+2}$ & 4.785 $10^{+3}$ & 4.495 $10^{+4}$ & 2.421 $10^{+5}$ & 6.721 $10^{+5}$ & 9.993 $10^{+5}$ & 3.120 $10^{+5}$ & 8.080 $10^{+4}$ & 2.824 $10^{+6}$ \\
\hline
\end{tabular}}
\caption{Vibrational transition probabilities (in s$^{-1}$) calculated by Brooke at al. \cite{BROOKE201311} for the d$^3\Pi_g$-a$^3\Pi_u$ transition. The transition probabilities are calculated as the sum of all the possible transition from a J'=1,$\Omega'$ =0 energy level.}
\label{tab:C2prob}
\end{table}

\section{Molecular constants for the calculations of the CN violet system}
\label{CNappendix}
\begin{table}[ht]
\centering
\begin{tabular}{|c|c|c|c|c|}
\hline
\multicolumn{5}{|c|}{$X^2\Sigma^+$}\\
\hline
$\nu$  & Origin & B$_\nu$ & $10^6 \cdot $ D$_\nu$ &$10^{-3}\cdot \gamma$ \\
\hline
0&	0.0			&1.891089596(96)&	6.39726(64)&	7.25514(52) \\
1&	2042.42143(24)	&1.873665288(90)	&6.40576(60)&	7.17376(74) \\
2&	4058.54933(29)&	1.856186883(85)&	6.41672(60)	&7.0850(12) \\
3&	6048.34329(35)&	1.83865221(11)	&6.42731(56)&	6.9814(12)\\
4&	8011.76637(42)&	1.82105914(21)&	6.44121(73)	&6.8631(14)\\
5&	9948.77554(56)&	1.80340409(27)&	6.4530(38)	&6.7198(14)\\
6&	11859.32721(61)&	1.78568472(29)&	6.4651(44)&	6.5417(15) \\
7&	13743.37442(66)&	1.76789824(29)&	6.4812(46)	&6.3136(14) \\
8&	15600.86884(71)	&1.75004020(28)&	6.4835(64)	&6.0121(15) \\
9&	17431.75410(77)	&1.73210149(27)&	6.5334(85)	&5.6133(22) \\
\hline
\end{tabular}
\caption{Molecular constants (in cm$^{-1}$) for the $X^2\Sigma^+$ used in the calculations (in cm$^{-1}$).The uncertainties are given in parenthesis (i.e. one standard deviation to the last signiﬁcant digits of the constants).}
\label{tab:CNCONSTANTSI}
\end{table}

\begin{table}[ht]
\centering
\begin{tabular}{|c|c|c|c|c|}
\hline
\multicolumn{5}{|c|}{$B^2\Sigma^+$}\\
\hline
$\nu$  & Origin & B$_\nu$ & $10^6 \cdot $ D$_\nu$ &$10^{-2}\cdot \gamma$ \\
\hline
0&	25797.86825(43)&	1.9587413(13)&	0.660855(81)&	1.7154(52) \\
1&	27921.46650(55)&	1.9380444(45)&	0.67324(29)&	1.8162(82) \\
2&	30004.90632(77)&	1.916503(10)&	0.7021(27)&	1.840(13)\\
3&	32045.94678(73)&	1.894180(15)&	0.7105(60)&	2.453(16) \\
4&	34041.97036(68)&	1.8704809(66)&	0.7448(15)&	2.1169(97) \\
5&	35990.0970(21)&	1.847108(24)&	0.9132(54)&	0.431(83) \\
6&	37887.42418(74)&	1.8193429(54)&	0.8092(11)&	2.5237(87) \\
7&	39730.53401(80)&	1.790761(12)&	1.1054(58)&	0.6126(58)\\
8&	41516.64296(84)&	1.7621417(59)&	0.9040(13)&	3.4942(98)\\
\hline
\end{tabular}
\caption{Molecular constants (in cm$^{-1}$) for the $B^2\Sigma^+$ used in the calculations (in cm$^{-1}$). The uncertainties are given in parenthesis (i.e. one standard deviation to the last signiﬁcant digits of the constants).}
\label{tab:CNCONSTANTSII}
\end{table}

\begin{table}[ht]
\resizebox{\textwidth}{!}{%
\begin{tabular}{|c|c|c|c|c|c|c|c|c|c|c|}
\hline

\multicolumn{2}{|c|}{\multirow{2}{*}{\rotatebox{25}{$\nu''-\nu'$}}}& \multicolumn{9}{c|}{$B^2\Sigma^+$}\\
\cline{3-11}

\multicolumn{2}{|c|}{}    &0      &1      &2       &3      &4    &5     &6    &7    &8      \\
\hline
\parbox[t]{2mm}{\multirow{8}{*}{\rotatebox[origin=c]{90}{$X^2\Sigma^+$}}}
&0 &4.11$10^{-2}$ & 3.80$10^{-3}$&	4.02$10^{-5}$&	1.06$10^{-6}$&	4.39$10^{-8}$&	2.27$10^{-9}$&	1.16$10^{-11}$& 1.22$10^{-11}$& 2.94$10^{-15}$\\
&1 &3.67$10^{-3}$ & 3.36$10^{-2}$&	6.50$10^{-3}$&	7.83$10^{-5}$&	5.17$10^{-6}$&	1.39$10^{-7}$&	1.58$10^{-8}$&	2.11$10^{-12}$& 5.35$10^{-11}$\\
&2 &2.47$10^{-4}$ & 6.04$10^{-3}$&	2.80$10^{-2}$&	8.38$10^{-3}$&	9.43$10^{-5}$&	1.49$10^{-5}$&	2.34$10^{-7}$&	5.98$10^{-8}$&	2.31$10^{-10}$\\
&3 &1.21$10^{-5}$ & 6.17$10^{-4}$& 7.50$10^{-3}$&	2.37$10^{-2}$&	9.64$10^{-3}$&	8.28$10^{-5}$&	3.27$10^{-5}$&	2.32$10^{-7}$&	1.74$10^{-7}$\\
&4 &3.21$10^{-7}$ & 4.14$10^{-5}$& 1.03$10^{-3}$&	8.32$10^{-3}$&	2.04$10^{-2}$&	1.04$10^{-2}$&	5.08$10^{-5}$&	6.02$10^{-5}$&	7.89$10^{-8}$\\
&5 &1.91$10^{-10} $& 1.50$10^{-6}$& 8.81$10^{-5}$&	1.43$10^{-3}$&	8.68$10^{-3}$&	1.80$10^{-2}$&	1.08$10^{-2}$&	1.54$10^{-5}$&	9.67$10^{-5}$\\
&6 &1.66$10^{-9} $& 4.00$10^{-9}$& 4.15$10^{-6}$&	1.50$10^{-4}$&	1.79$10^{-3}$&	8.72$10^{-3}$&	1.62$10^{-2}$&	1.08$10^{-2}$&	1.80$10^{-7}$\\
&7 &5.40$10^{-10} $&7.21$10^{-9}$& 2.72$10^{-8}$&	8.79$10^{-6}$&	2.22$10^{-4}$&	2.09$10^{-3}$&	8.57$10^{-3}$&	1.48$10^{-2}$&	1.05$10^{-2}$\\
&8 &7.23$10^{-11} $&2.89$10^{-9}$& 1.72$10^{-8}$&	1.12$10^{-7}$&	1.57$10^{-5}$&	3.01$10^{-4}$&	2.32$10^{-3}$&	8.26$10^{-3}$& 1.38$10^{-2}$\\
&9 &2.76$10^{-12} $&4.39$10^{-10}$& 8.65$10^{-9}$& 2.75$10^{-8}$& 3.23$10^{-7}$& 2.53$10^{-5}$& 3.82$10^{-4}$& 2.50$10^{-3}$&	7.85$10^{-3}$\\
\hline
\end{tabular}}
\caption{The table shows the transition probabilities extracted from LIFBASE for the CN (B$^2\Sigma^+$-X$^2\Sigma^+$) transition.}
\label{tab:CNTDM}
\end{table}

\end{appendices}
\end{document}